\documentstyle[editedvolume,epsfig]{crckapb} 

\def\ltsim{\raise 2pt \hbox {$<$} \kern-1.1em \lower 4pt \hbox {$\sim$}}

\begin{opening}
\title{NON--THERMAL EMISSION FROM EXTRAGALACTIC RADIO SOURCES:
A HIGH RESOLUTION -- BROAD BAND APPROACH}

\author{GIANFRANCO BRUNETTI}
\institute{\it Istituto di Radioastronomia del CNR\\
           via P. Gobetti 101, I-40129 Bologna, Italy}

\end{opening}

\runningtitle{Spatial resolution in X--ray astronomy}

\begin{document}

     
\section{ABSTRACT}

In the framework of the study of
extragalactic radio
sources, we will focus on the importance of the
spatial resolution at different wavelengths, and 
of the combination of observations at different frequency
bands.
In particular, a substantial step forward in this field
is now provided by the new generation X-ray telescopes 
which are able to image radio 
sources in between 0.1--10 keV with a spatial 
resolution comparable with that of the radio telescopes
(VLA) and of the optical telescopes.
After a brief description 
of some basic aspects of acceleration mechanisms 
and of the 
radiative processes at work in the extragalactic 
radio sources, we will focus on a number of recent  
radio, optical and X--ray observations with arcsec resolution,
and discuss the deriving constraints on the 
physics of these sources.

\section{INTRODUCTION}

This contribution is focussed on some basic concepts regarding
the study of the non--thermal emission from 
extragalactic radio sources based on a 
broad band, multifrequency approach.
The principal difficulty in this study arises by the
different sensitivity and spatial resolution of the 
instruments in different bands.
Radio telescopes easily reach sub--arcsec spatial resolutions and 
can image very faint sources by relatively short 
observations. 
On the other hand, optical telescopes generally have 
only arcsec spatial resolution so that 
combined radio -- optical studies are limited
by the lower spatial resolution of the optical telescopes.
Although, the problem is alleviated 
by making use of optical HST observations, it still remains
to a certain degree.
In order to extend this approach to higher energies,
X--ray telescopes with arcsec resolution (at least)
and 
good sensitivity are required.
Neverthless the poor spatial resolution of the
past X--ray observatories, 
pioneeristic studies in this direction have been 
attempted in the last 10-20 years with combined radio 
(and optical) and {\it Einstein} or ROSAT 
X--ray observations.
\begin{table}[htb]
\begin{center}
\caption{X-ray Observatories}
\begin{tabular}{lllllll}
\hline
\hline
Satellite & Instrument & Flux$\rightarrow$1 cts/s & Energy Band & 
Resolution \\
     &     & (erg/s/cm$^{2}$) & (keV)  & (arcsec) \\
\hline
Einstein & HRI & 1.6E-10 & 0.5-4.0   &   4         \\
         & IPC & 2.8E-11 & 0.5-4.0   &             \\
\hline
ROSAT    & HRI & 3.7E-11 & 0.5-2.4 &   3*         \\
         & PSPC& 1.4E-11 & 0.5-2.4 &  20         \\
\hline
ASCA     & SIS & 3.3E-11 & 0.4-12  & 180         \\
         & GIS & 3.8E-11 & 0.4-12  & 180         \\
\hline
BeppoSAX & MECS& 8.1E-11 & 1.3-10  &    60         \\
	 & PDS & 9.5E-11   & 13-80 &             \\
\hline
Chandra  & HRC-I& 2.8E-11  & 0.4-10& 0.5         \\
         & ACIS-I& 1.1E-11 & 0.4-10& 0.5         \\
\hline
\hline
\end{tabular}

{\bf Notes}: {\bf Column 3},  
Flux$\rightarrow$ is the flux (erg/s/cm$^2$) necessary
to have 1 cts/s in the detector.
{\bf Column 5}: (*) means that the spatial resolution 
results affected by errors in the aspect solution 
associated with the wobble of the space craft. 
\end{center}
\end{table}
More recently,
a substantial step forward has been achieved thanks to the
advent of the new generation X--ray satellites:
{\it Chandra} and XMM-{\it Newton}.
In Tab.1 we report
the main capabilities of a selection of
past X--ray observatories compared
with those of {\it Chandra}:
the abrupt increase of the spatial resolution combined with
the high effective area of {\it Chandra}
represent a `new revolution' in astrophysics
(data taken from Cox, 1999).
For the first time it is now possible to investigate
the X--ray properties of extragalactic radio sources
performing spatially resolved spectroscopy over a relatively
large energy band (about 0.4--10 keV).
{\it Chandra} allows us to
obtain images with $\leq$arcsec spatial resolution, comparable
with the typical resolution of the VLA radio images and
of the optical telescopes and thus
it is a tool to perform, for the first time,
multiwavelength studies from the radio band to the X--rays.

As we will show in this contribution,
the high resolution broad band study of
non--thermal emission from the extragalactic radio sources
allows us to sample the emission due to very different
portions of the spectrum of the relativistic electrons or to 
study the emission due to different emitting mechanisms at work
in the observed regions.
After the first 3 years of new {\it Chandra} observations,
it is now clear that
the scenarios describing 
these objects should now be partially revisited.

In Sect.3 we discuss some theoretical aspects related to 
the relativistic plasma in extragalactic radio sources.
More specifically, in Sects. 3.1--3.3 we report the 
basic concepts of particle acceleration and discuss 
the resulting spectrum of the relativistic electrons in the
framework of the standard shock acceleration scenario.
In Sect. 3.4 we introduce the non--thermal
emitting mechanisms. In particular, we focus
on the sampling of the spectrum of the emitting
electrons provided
by the observations at different frequencies.
In Sect. 3.5 we review the basic methods to derive
the energetics of the relativistic plasma in the
extragalactic radio sources.
Finally, in Sect.4 we concentrate on the most recent high resolution 
multiwavelength studies of extragalactic radio sources.
Our `biased' 
review is mainly focussed on the new {\it Chandra} observations
of radio lobes, radio jets and radio hot spots.
Combining the data with the methods showed 
in Sect. 3, 
we discuss on the new constraints on the shape of
the low and high energy end of the electron spectrum,
on the kinematics of radio jets, and on the
energetics of extragalactic radio sources.

\section{PARTICLE ACCELERATION AND RADIATIVE MECHANISMS}

In this Section we give the basic concepts
of the particle acceleration and of the time
evolution of the energy of the particles. 
We focus our attention on the case 
of electrons which are responsible 
for the observed non--thermal emission.
The acceleration of electrons to high energies in various
environments is a problem of widespread interest
in astrophysics.
Indeed, the synchrotron emission of jets in a large number of 
extragalactic radio sources (e.g., Miley, 1980; Fomalont, 1983;
Venturi, this proceedings) requires in situ acceleration
of electrons because of the short
synchrotron life--times (e.g., Rees, 1978; 
Bicknell \& Melrose, 1982).
Also the radio luminosity of supernova remnants and their broad
band spectrum (e.g., Clark \& Caswell, 1976; Dickel, 1983) 
suggest that electrons are accelerated to high energy 
in these cases.
Finally, the diffuse Mpc--size radio emission discovered in 
an increasing number of clusters of galaxies 
(e.g., Feretti \& Giovannini 1996) requires large scale in situ 
acceleration of protons and electrons 
(e.g., Jaffe 1977; Tribble 1993;
Sarazin 1999; Blasi 2001; Brunetti et al. 2001a; Petrosian 2001).

The basic goal of this Section is simply  
to show, based on 
general theoretical arguments, that 
the energy distribution of the 
relativistic electrons in extragalactic radio sources
cannot be a simple power law.
Although a power law energy distribution is commonly assumed 
in the literature, here we will point out that
this approximation cannot be adopted 
in a broad band, multiwavelength (radio to X--ray observations)
approach, and further
that the observations can be used to constrain
the energy distribution of the emitting electrons.
In this way we will test the acceleration theories.

\subsection{Competing mechanisms at work}

The relativistic electrons in extragalactic radio sources
are subject to several inescapable
processes that change their energy
with time.
For a general description of the leading mechanisms
at work we refer the reader to classical books
(e.g., Ginzburg 1969; Pacholczyk 1970; Rybicki \& Lightman 1979;
Melrose 1980).

Here, it is important to underline that the efficiency of
these processes is related to the energy of the electrons
in a way that depends on the particular process.
As a consequence, the time evolution
of relativistic electrons at different energies
is expected to be dominated by different processes.
In the following, we give some basic relationships:

$\bullet$
Relativistic electrons with Lorentz factor $\gamma$
mixed with a thermal plasma
lose energy via ionization losses and Coulomb collisions;
for relativistic electrons one has :

\begin{equation}
\left( {{ d \gamma }\over{d t}}\right)_{\rm ion}^-=-
1.2 \times 10^{-12} n 
\left[1+ {{ {\rm ln}(\gamma/n) }\over{
75 }} \right]
\label{ion}
\end{equation}

\noindent
where $n$ is the number density of the thermal plasma.

$\bullet$
If the relativistic electrons are confined into an
expanding region, they lose energy via adiabatic
expansion:

\begin{equation}
\left( {{ d \gamma }\over{d t}}\right)_{\rm ex}^-=-
{1\over{R}} {{ d R}\over{dt}} \gamma
\sim 10^{-11} \gamma {{ v_{\rm ex}(t)}\over{c}}
{1\over{R(t)_{\rm kpc}}}
\label{ad}
\end{equation}

\noindent
where 
$v_{\rm ex}$ and $R$ are the expansion velocity and dimension
of the region, respectively.
 
$\bullet$
Relativistic electrons in a magnetic field $B$ lose energy
via synchrotron radiation :

\begin{equation}
\left( {{ d \gamma }\over{d t}}\right)_{\rm syn}^-=-
1.9 \times 10^{-9} \gamma^2 B^2 \sin^2\theta
\label{syn}
\end{equation}

\noindent
and via inverse Compton scattering with photons (of
energy density $\omega_{\rm ph}$); in the Thompson
regime, one has:

\begin{equation}
\left( {{ d \gamma }\over{d t}} \right)_{\rm ic}^-=-
3.2 \times 10^{-8} \gamma^2 \omega_{\rm ph}
\label{ic}
\end{equation}

$\bullet$
Finally, the electrons can be accelerated/re--accelerated
by several mechanisms. The most commonly considered
in the literature are the Fermi and Fermi--like
processes. It should be stressed that, in general, 
such mechanisms are stochastic processes, i.e., 
given an initial monoenergetic electron distribution,
they not only produce a net increase of the energy of the 
electrons but also a broadening of the energy distribution. 
Thus the effect of a reacceleration mechanism
on an initial energy distribution can be described by
the combination of a systematic acceleration of the electrons
and of a 'statistical' broadening of the resulting
energy distribution. 
   
\noindent   
Two relevant cases are:
 
{\it a) acceleration due to MHD
turbulence}; this is discussed in detail in a number of papers and
classical
books (e.g., Melrose 1980).
If the resonance scattering condition (e.g., Eilek \& Hughes 1991)
is satisfied (this in general requires already
relativistic electrons, e.g. Hamilton \& Petrosian 1992), 
turbulent Alfven waves can accelerate electrons via resonant
pitch angle scattering.
For a power law energy spectrum of the Alfven waves: 

\begin{equation}
P(k)=
b {{ B^2}\over{8 \pi}}
{{ s-1}\over{k_o}} 
\left(
{{ k \over {k_o} }}
\right)^{-s}
\label{aspectrum}
\end{equation}

in the range $k_o < k < k_{\rm max}$, where
$k$ is the wave number 
($k_o << k_{\rm max}$) and $b$ is
a normalization factor indicating the
fractional energy density in waves, the systematic energy gain is
(e.g., Isenberg 1987; Blasi 2000; Ohno et al. 2002):

\begin{equation}
\left( {{ d \gamma }\over{d t}}\right)_{\rm A-tur}^+
\sim 
(s+2) {\xi_{\rm A} }
\gamma^{s-1}
\label{ono1}
\end{equation}

where

\begin{equation}
{\xi_{\rm A} }=
{{s-1}\over{s(s+2)}}
{{ \pi b k_o {v_{\rm A}}^2}\over{
c}}
\left(
{{ e B }\over{m_{\rm e} c^2 k_o}}
\right)^{2-s}
\label{ono2}
\end{equation}

and where $v_{\rm A}$ is the Alfven velocity.

\noindent
MHD turbulence can also accelerate relativistic particles
in radio sources 
via Fermi--like processes (e.g., Lacombe 1977; Ferrari et al. 1979).
Under the simple assumption of a quasi--monochromatic
turbulent scale (e.g., Gitti et al. 2002) 
the systematic energy gain is :

\begin{equation}
\left( {{ d \gamma }\over{d t}}\right)_{\rm F-tur}^+
\simeq 4 \times 10^{-11} \gamma
{{ {v_{\rm A}}^2 }\over{ l }}
\left( {{\delta B}\over{B}} \right)^2
\label{tur}
\end{equation}

where $l$ is the
distance between two peaks of turbulence and 
$\delta B/B$ is the fluctuation in a peak 
of the field intensity with
respect to the average field strength.

{\it b) acceleration in a shock}; (e.g., Meisenheimer et al. 1989);
one has :

\begin{equation}
\left( {{ d \gamma }\over{d t}}\right)_{\rm sh}^+
\simeq \gamma {{ {U_-}^2 c }\over{r}} \left( {{ r-1}\over{
r\lambda_+ + \lambda_- }} \right)
\label{sho}
\end{equation}

where $U_-$ is the velocity of the plasma in the
region
before the shock discontinuity (measured in the
shock frame and in unit of $c$), $r$ is the
shock compression ratio, and $\lambda_{\pm}$ is
the mean diffusion length of the electrons in the region 
before ($-$) and after ($+$) the shock.
Assuming a similar $B$--field configuration and strength before
and after the shock (i.e. at some distance from the shock in the
pre-- and post--shock region) Eq.(\ref{sho}) can be generalized
as :

\begin{equation}
\left( {{ d \gamma }\over{d t}}\right)_{\rm sh}^+
\simeq \gamma {{ {U_-}^2 }\over{r}} \left( {{ r-1}\over{ r+1}}
\right) {1 \over {3 {\cal K}(\gamma) }}
\label{shok}
\end{equation}

\noindent
where 
${\cal K}(\gamma)$ is the spatial diffusion coefficient which
depends on the physical conditions.

In the relevant astrophysical cases the time evolution 
of the energy of the electrons is given by the
combination of several mechanisms, i.e. :

\begin{equation}
\left( {{ d \gamma }\over{d t}}\right)=
\sum_i^{..} \left( {{ d \gamma }\over{d t}}\right)_i^-+
\sum_i^{..} \left( {{ d \gamma }\over{d t}}\right)_i^+
\end{equation}

\begin{figure}
\resizebox{\hsize}{!}{\includegraphics{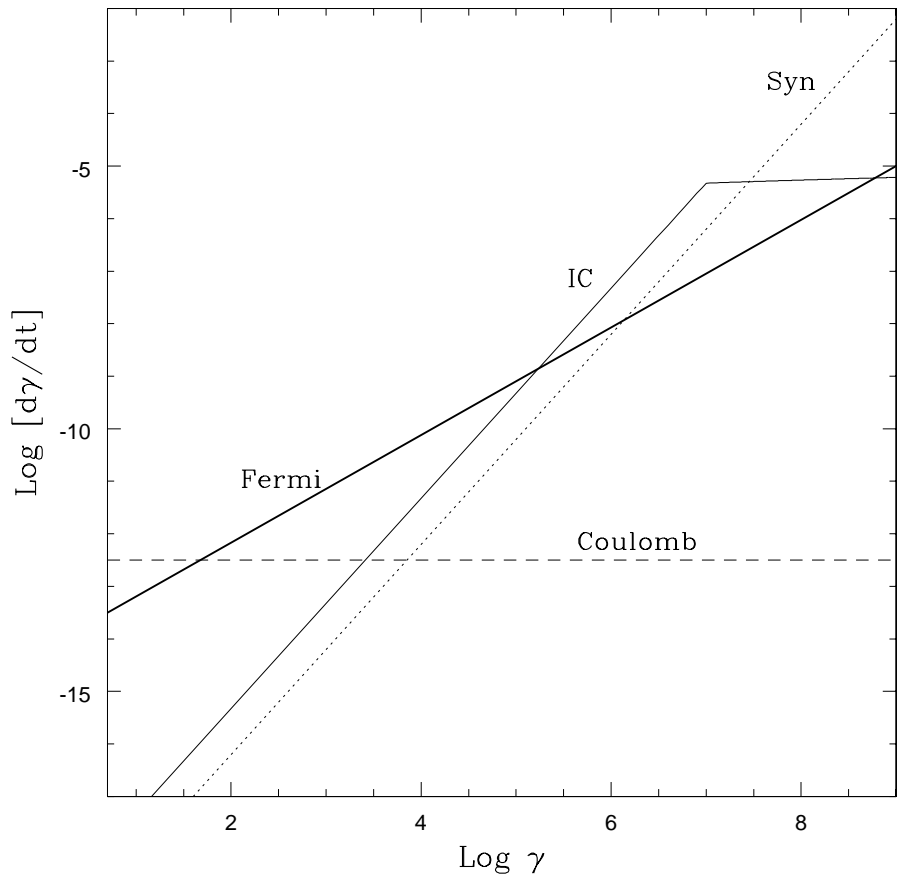}}
\caption{A sketch of the efficiency of acceleration (solid line) and
loss (dashed lines) processes is shown, as a function of
the Lorentz factor of the electrons.
The artificial knee introduced in the IC losses at $\gamma
>10^7$ simulates the effect of the KN cross section.}
\end{figure}

In Fig.1 we report $d\gamma/dt$ due to the
above processes as a function of $\gamma$
for representative physical conditions.
It is clear that the different processes dominate in different
energy regions with the Coulomb losses being especially important
at low energies.
In addition, due to its dependence on the
energy ($\propto \gamma$), the Fermi--like acceleration
can be an efficient process in a restricted energy band, whereas
Coulomb and radiative losses in general, prevent electron
acceleration at lower and higher energies, respectively.

\subsection{The basic equations of particle evolution}

In order to consider the general 
evolution of the particle spectrum, a kinetic theory approach is 
suitable (e.g., Melrose 1980; 
Livshitz \& Pitaevskii 1981; Blandford 1986;
Blandford \& Eichler 1987; 
Eilek \& Hughes 1991, and reference therein).
Given $f({\bf p},..)$, the distribution function of the electrons 
(i.e. $f({\bf p},..) d{\bf p}$ is the number of electrons in the 
element $d{\bf p}$ of the momentum space), 
the description of the behaviour of this distribution function
is given by a Boltzman equation:

\begin{equation}
{{d f({\bf p})}\over{dt}}=
\left( {{\partial f}\over{
\partial t}} \right)_{\rm coll}
+
\left( {{\partial f}\over{
\partial t}} \right)_{\rm diff}
\label{eilek}
\end{equation}

\noindent
where, the total
time derivative is to be interpreted according to:

\begin{equation}
{{d }\over{dt}} \rightarrow
{{\partial}\over{\partial t}}
+ {\bf v} \cdot {{\partial}\over{\partial {\bf r}}}
+ {\bf F} \cdot {{\partial}\over{\partial {\bf p}}}
\end{equation}

\noindent
where ${\bf v}$ is the particle velocity, 
and ${\bf F}=d{\bf p}/dt$
the force acting on the particles.

The {\it diffusion} term in Eq.(\ref{eilek}) describes spatial diffusion,
while the {\it collision} term accounts for all the physics of
collisions and scattering (e.g., 
radiative losses, interaction with waves and shocks, 
Coulomb collisions).
A general, stochastic acceleration may be thought of as a diffusion
in the momentum space given by a diffusion coefficient, $D_{pp}$, so
that in case of no spatial diffusion and no losses, 
Eq.(\ref{eilek}) can be
written as:

\begin{equation}
{{d f({\bf p})}\over{dt}}=
{{\partial}\over{\partial p_i}}
[ (D_{pp})_{ij}({\bf p}) {{\partial}\over{\partial p_i}}
f({\bf p})]
\end{equation}

\noindent
If isotropy of the scattering waves and electrons is assumed, 
$f({\bf p})d{\bf p}= 4 \pi p^2 f(p) dp$ and Eq.(\ref{eilek})
can be written as (e.g., Melrose 1980; Eilek \& Hughes 1991):

\begin{eqnarray}
{{\partial }\over{\partial t}} f({\bf r},p,t)
+{{\partial }\over{\partial x_i}}
[{\cal K}_{ij} {{\partial }\over{\partial x_j}} f({\bf r},p,t) ]
=
{1\over{p^2}} {{\partial }\over{\partial p}}
[{\cal S} p^4 f({\bf r},p,t) \nonumber\\
+p^2 D_{pp} {{\partial }\over{\partial p}} f({\bf r},p,t)+
p^2 I(p) f({\bf r},p,t)] + Q({\bf r},p,t)
\label{continuita1}
\end{eqnarray}

\noindent 
where we have considered the effect of the synchrotron and inverse
Compton losses ($dp/dt=- {\cal S} p^2$), of the
Coulomb losses ($dp/dt=- I(p)$) and 
considered a general form for the spatial diffusion of the
electrons (${\cal K}$ being the spatial diffusion coefficient).
A term accounting for electron injection (injection rate = $Q$)
has also been introduced.

\noindent
Eq.(\ref{continuita1}) can be also written in the energy space.
For an isotropic distribution of the momenta of the 
relativistic electrons, the electron spectrum is
$N(\epsilon)=4\pi p^2 f(p) dp/d\epsilon$ and one has :

\begin{equation}
{{\partial }\over{\partial t}} f({\bf r},p,t)
= {{\rm c}\over{ 4 \pi p^2}} {{\partial }\over{\partial t}} 
N({\bf r},\epsilon,t)
\label{continuita1_1}
\end{equation}

\begin{equation}
{{\partial }\over{\partial p}} [{\cal S} p^4 f({\bf r},p,t)]
= {{\rm c}\over{ 4 \pi }} {{\partial }\over{\partial p}} 
[{\cal S} p^2 N({\bf r},\epsilon,t)]
\label{continuita1_2}
\end{equation}

\begin{equation}
{{\partial }\over{\partial p}} [p^2 I(p) f({\bf r},p,t)]
= {{\rm c}\over{ 4 \pi }} {{\partial }\over{\partial p}} 
[I(p) N({\bf r},\epsilon,t)]
\label{continuita1_3}
\end{equation}

\begin{equation}
{{\partial }\over{\partial p}} [p^2 D_{pp}
{{\partial }\over{\partial p}} f({\bf r},p,t)]=
{{\rm c}\over{ 4 \pi }} {{\partial }\over{\partial p}}
[D_{pp} {{\partial }\over{\partial p}} N({\bf r},\epsilon,t)
-{2\over{p}} D_{pp} N({\bf r},\epsilon,t)]
\label{continuita1_4}
\end{equation}

\noindent
thus Eq.(\ref{continuita1}) can be written as:

\begin{eqnarray}
{{\partial }\over{\partial t}} N({\bf r},\epsilon,t)
+{{\partial }\over{\partial x_i}}
[{\cal K}_{ij} {{\partial }\over{\partial x_j}} 
N({\bf r},\epsilon,t) ]
= {{\partial }\over{\partial p}}
[{\cal S} p^2 N({\bf r},\epsilon,t) \nonumber\\
I(p) N({\bf r},\epsilon,t) +
+ D_{pp} {{\partial }\over{\partial p}} N({\bf r},\epsilon,t)
- {2\over{p}} D_{pp} N({\bf r},p,t)] + Q(x,\epsilon,t)
\label{continuita1_new}
\end{eqnarray}

\noindent
Let us assume that the reacceleration processes are Fermi--like, 
i.e. $D_{pp}={{\chi}\over{2}} p^2$ and that the Coulomb losses are
negligible, i.e. $I(p) \sim 0$.
In terms of Lorentz factors and neglecting the dependence on the
spatial coordinates, it is possible to rewrite 
Eq.(\ref{continuita1_new}) as :

\begin{equation}
{{\partial }\over{\partial t}} N(\gamma,t)=
-{{\partial }\over{\partial \gamma}} 
\left[{{ d \gamma}\over{d t}} N(\gamma, t) \right]+
{{\chi}\over 2} {{ \partial }\over{\partial \gamma}}\left[
\gamma^2 {{ \partial N(\gamma,t)}\over{\partial \gamma}}
\right]
-{{ N(\gamma,t)}\over{T_{\rm es}}}
+Q(\gamma,t)
\label{continuita2}
\end{equation}

where

\begin{equation}
{{d \gamma}\over{d t}}=
\chi \gamma - {\cal S} \gamma^2 {\rm mc}
\end{equation}

for simplicity we have replaced the general spatial
diffusion term with a simple term $N/T_{\rm es}$, accounting for
an energy independent diffusion of the electrons from the system. 
For a general treatment of the solution of Eq.(\ref{continuita2})
we refer the reader to Kardashev (1962). 
Some additional more recent applications can be found in a number 
of papers
(e.g., Borovsky \& Eilek 1986; 
Sarazin 1999; Brunetti et al. 2001a; Petrosian 2001 and
references therein).

\subsection{Shock acceleration and electron spectrum}

Shock acceleration is a well known astrophysical process
(see Drury 1983 and Blandford \& Eichler 1987 for
a review).
Energetic charged particles can be efficiently 
accelerated in shock waves either by drifts
in the electric fields at the shock (e.g., Webb et al., 
1983 and references therein), or by scattering back and
forth across the shock in the magnetic turbulence present in the
background plasma (e.g., Bell 1978a,b; Blandford 1979;
Drury 1983).
In the simplest case, i.e. that of a steady state shock
without losses and with a seed monoenergetic particle distribution, 
the accelerated spectrum is a simple power law
(e.g., Bell 1978a,b; Blandford \& Ostriker, 1978).
On the other hand, the effect of synchrotron 
and inverse Compton
losses in diffusive shock acceleration causes
the development of a cut--off in the accelerated
spectrum and a break in the electron spectrum
integrated over the post shock region 
(e.g., Meisenheimer et al. 1989). 
Detailed studies of the diffusive shock acceleration
including energy losses under steady state conditions, 
find the development of both a high energy cut--off, 
and humps in the
accelerated spectrum which depend on the physical conditions
(Webb et al. 1984).
Finally, more recent studies have investigated non steady state  
diffusive shock acceleration (Webb \& Fritz 1990),
multiple shocks acceleration (e.g., Micono et al. 1999;
Marcowith \& Kirk 1999), and 
particle acceleration via relativistic shocks
(e.g., Kirk et al. 2000).

\subsubsection{A simplified semi-analytic approach}

In this Section we discuss a relatively simple  
analytic method to derive the energy
distribution of the relativistic electrons accelerated
in a shock region.
In what follows we assumed that :

a) relativistic electrons, continuously injected in the
shock region with a power law energy distribution are 
(re-)accelerated in this
region subject to radiative losses.
The final spectrum
in the shock region 
is thus given by the competition between acceleration and 
loss terms;

b) the accelerated electrons diffuse from
the shock region and continuously fill a post--shock
region in which most of the observed emission is produced.
As an additional simplification, we assume that 
the electrons are transported throughout the post--shock region with
a constant velocity so that older electrons are located
at larger distances from the shock.
In the post--shock region the electrons are not (re-)accelerated
but they are only subjects to radiative losses so that
the energy of the electrons decreases with time.

As in Kirk et al.(1998), the electrons
are assumed to be 
reaccelerated by Fermi--I like processes in the shock
region from which they typically escape in a time
$T_{\rm es}$.
To calculate the electron spectrum in the shock region
we can solve Eq.(\ref{continuita2}) with a time independent 
approach (i.e. $\partial N / \partial t =0$):

\begin{equation}
\left({{ d}\over{d \gamma}} N(\gamma) \right) {{d\gamma}\over
{dt}} + N(\gamma)
\left[ {d\over d\gamma} {{d\gamma}\over{dt}} +
{1\over{T_{\rm es}}} \right]
-Q(\gamma)=0
\label{tind}
\end{equation}

\noindent
where

\begin{equation}
{{d \gamma }\over{dt}}=\chi \gamma - \beta \gamma^2
\label{dgdtshock}
\end{equation}

$\beta= {\rm mc} {\cal S}$ in Eq.(22), 
and $Q(\gamma)=Q \gamma^{-s}$ (for $\gamma<\gamma_*$).
The solution is :

\begin{equation}
N(\gamma)=
{{ Q }\over{\chi}}
\left( 1- {{\gamma}\over{\gamma_{\rm c}}} \right)^{-\alpha_{-}}
\gamma^{-\alpha_{+}}
\int_{\gamma_{\rm low}}^{ {\tilde \gamma} }
{{ y^{ \alpha_{+} -(1 +s) } dy }\over{
[ 1 - y/\gamma_{\rm c} ]^{1-\alpha_-} }}
\label{nshock1}
\end{equation}

where $\alpha_{\pm}=1 \pm [\chi T_{\rm es} ]^{-1}$, 
${\tilde \gamma}={\rm min}(\gamma_*,\gamma)$, and
$\gamma_{\rm low}$ is an {\it artificial} cut--off
introduced in the integration of the equations corresponding
to the minimum energy of the electrons which can be
accelerated by the shock.
Indeed, it is well known that only particles with a Larmor radius
larger than the thickness of the shock are actually able to
`feel' the discontinuity at the shock (e.g., Eilek \& Hughes
1991 and ref. therein).
The shock thickness is of the order of the Larmor radius of
thermal protons so that, as a first approximation,  
we might use $\gamma_{\rm low} \geq 10$ in the case of electrons.

\begin{figure}
\resizebox{\hsize}{!}{\includegraphics{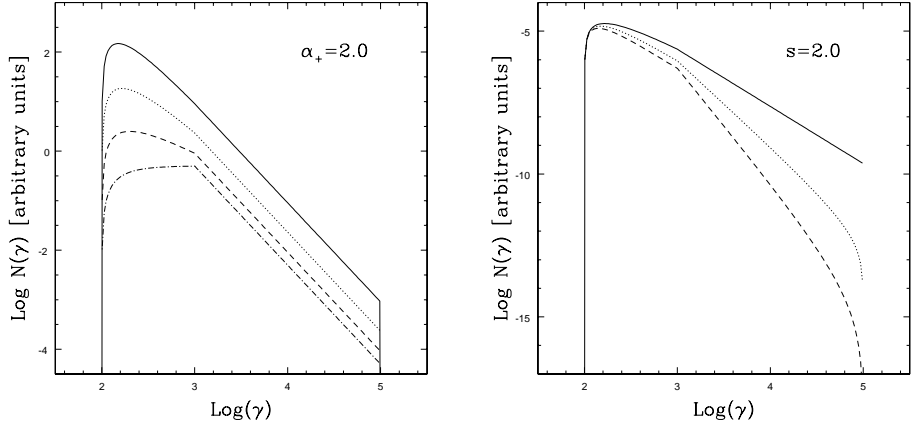}}
\caption{The time independent 
electron spectra in the shock region are reported
as a function of the energy of the electrons for 
different values of the relevant parameters.
Left panel: from the bottom we have:
$s$=0, 1, 2, 3; $\alpha_+=2.0$.
Right panel: from the bottom we have: $\alpha_+$=4, 3, 2;
$s=2.0$.}
\end{figure}

The electron spectrum in the shock region is reported in Fig.2:
a low energy cut--off is formed around $\gamma_{\rm low}$, whereas
a high energy cut--off is formed at $\gamma_{\rm c}=\chi/\beta$,
i.e. the energy at which the radiative losses outweight the acceleration
efficiency. An additional break in the spectrum is obtained around
$\gamma_*$; this break is due to the fact that, for $\gamma > \gamma_*$,
all the electrons injected in the shock region can be accelerated
at higher energies, whereas for $\gamma < \gamma_*$, only the electrons
injected with energy $<\gamma$ can contribute to the accelerated
spectrum.

As already anticipated, during the step b), once the electrons 
are re--accelerated
in the shock region, they travel towards the post shock region 
subject to the radiative losses only. The evolution of the 
spectrum (\ref{nshock1}) is obtained solving the {\it continuity}
equation (Eq.\ref{continuita2}) taking into account
only the effect of the radiative losses (i.e.,
$d\gamma / dt = -\beta \gamma^2$) :

\begin{equation}
{{ \partial N(\gamma,t) }\over{
\partial t}}=
-{{ \partial }\over{ \partial \gamma}}
\left[{{d \gamma}\over{dt}} N(\gamma,t) \right]
\label{nt}
\end{equation}

\noindent
assuming an approximately constant magnetic field strength
(i.e., $\beta =$const.), and a constant diffusion velocity 
$v_{\rm D}$, the resulting spectrum of the electrons
at a distance $r=v_{\rm D} t$ is :

\begin{eqnarray}
N(\gamma,t)= {{ Q }\over{\chi}} (1- \beta \gamma t)^{\alpha_+ -2}
\left[ 1 - {{ \gamma / \gamma_{\rm c} }\over{
1 - \beta \gamma t }} \right]^{-\alpha_-}
\gamma^{-\alpha_+} \times \nonumber\\
\int_{\gamma_{\rm low}}^{ {\tilde \gamma}(t) }
y^{\alpha_+ - (s+1) }
\left[ 1 - {{ y }\over{\gamma_{\rm c}}} \right]^{\alpha_--1}
\left[ 1 - {{ y }\over{\gamma_*}} \right]^{s-2} dy
\label{nshock2}
\end{eqnarray}

where 

\begin{equation}
{\tilde \gamma}(t)={\rm min}\left\{
\gamma_*, {{ \gamma }\over{1-\gamma \beta t}}
\right\}
\label{gmint}
\end{equation}

Fig.3 shows the time/spatial
evolution of the spectrum in the post--shock region.  

In order to further simplify the scenario, we might assume
that the spectrum of the injected electrons and that of the
accelerated electrons have the
same slope, i.e. $s=\alpha_+$ (as roughly
expected if the electrons
are continuously reaccelerated and released
by a series of similar shocks in the jets) and that the
acceleration time in the shock roughly equals the 
escape time, i.e.   
$\chi \sim {1\over{T_{\rm es}}}$.
In this case Eq.(\ref{nshock2}) becomes :

\begin{equation}
N(\gamma,t) \simeq Q \gamma^{-2} \int_{\gamma_{\rm low}}^
{ {\tilde \gamma}(t)} y^{-1} \left[
1 - {{ y }\over{ \gamma_{\rm c}}} \right]^{-1} dy
\label{nshock22}
\end{equation}

\begin{figure}
\resizebox{\hsize}{!}{\includegraphics{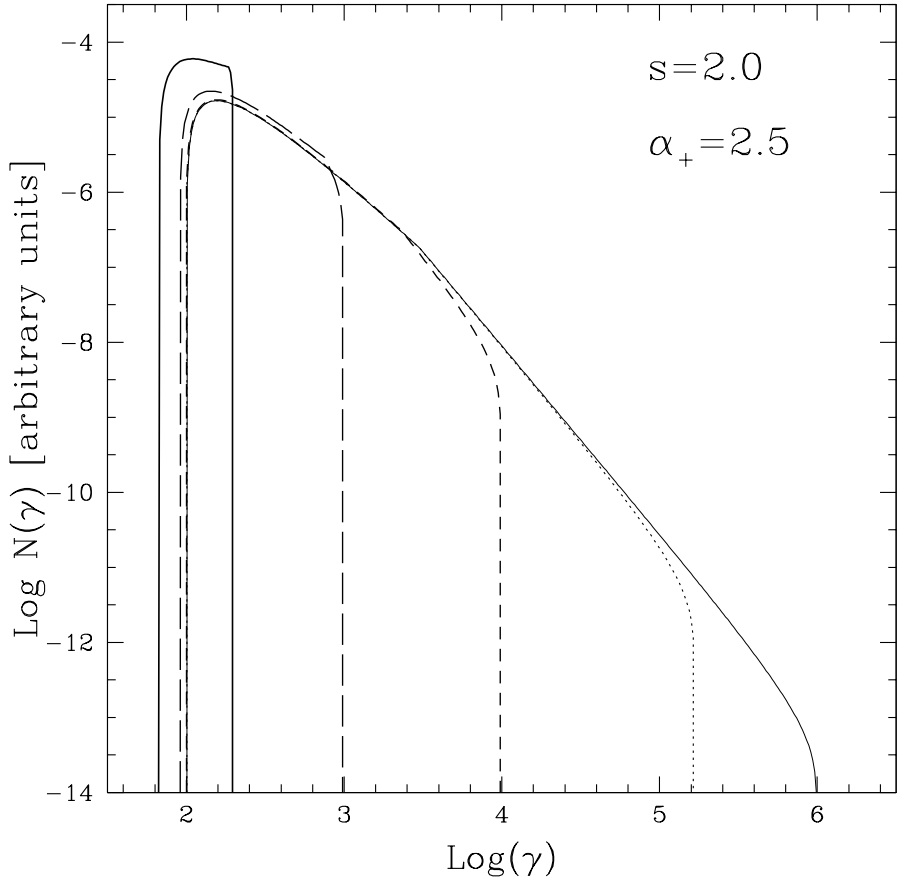}}
\caption{The time (or spatial) evolution 
of the electron spectrum in the post--shock region is
reported for $s=2.0$, $\alpha_+=2.5$, $\chi=10^{-12}$s$^{-1}$
and $\beta=10^{-18}$s$^{-1}$.
The different spectra correspond to a time after injection in the
post-shock region $\tau$= 0 (solid line), $5/ \chi$ (dotted line),
$10^2/ \chi$ (dashed line), $10^3 /\chi$ (long dashed line),
$5\times 10^3/ \chi$ (thick solid line).
}
\end{figure}

In general, the lack of spatial resolution better than 
1 arcsec over a wide energy band does not allow 
the resolution of the post--shock region in extragalactic radio sources
at different frequencies.
In fact, only information on the integrated emission from this
region can be obtained.
As a consequence, in order to obtain theoretical results
to be compared with observations,
one should deal with the electron spectrum
integrated over the post--shock region.
The size of this region, $L$, is determined
by the diffusion length of the electrons in the largest
considered time, $T_{\rm L}$, so that 
the volume integrated spectrum
of the electron population, $N_i(\gamma)$, is given by the
sum of all the electron spectra in this region, i.e. it is obtained
by integrating Eq(\ref{nshock2}, or \ref{nshock22})
over the time interval $0-T_{\rm L}$
(or in an equivalent way over the distance interval
from the shock $0-L$ taking into account the relationship
linking time and space). 
Assuming Eq.(\ref{nshock22}), the resulting integrated spectrum 
is reported in Fig.4 with the breaks and cut--offs 
indicated in the panel.

\begin{figure}
\resizebox{\hsize}{!}{\includegraphics{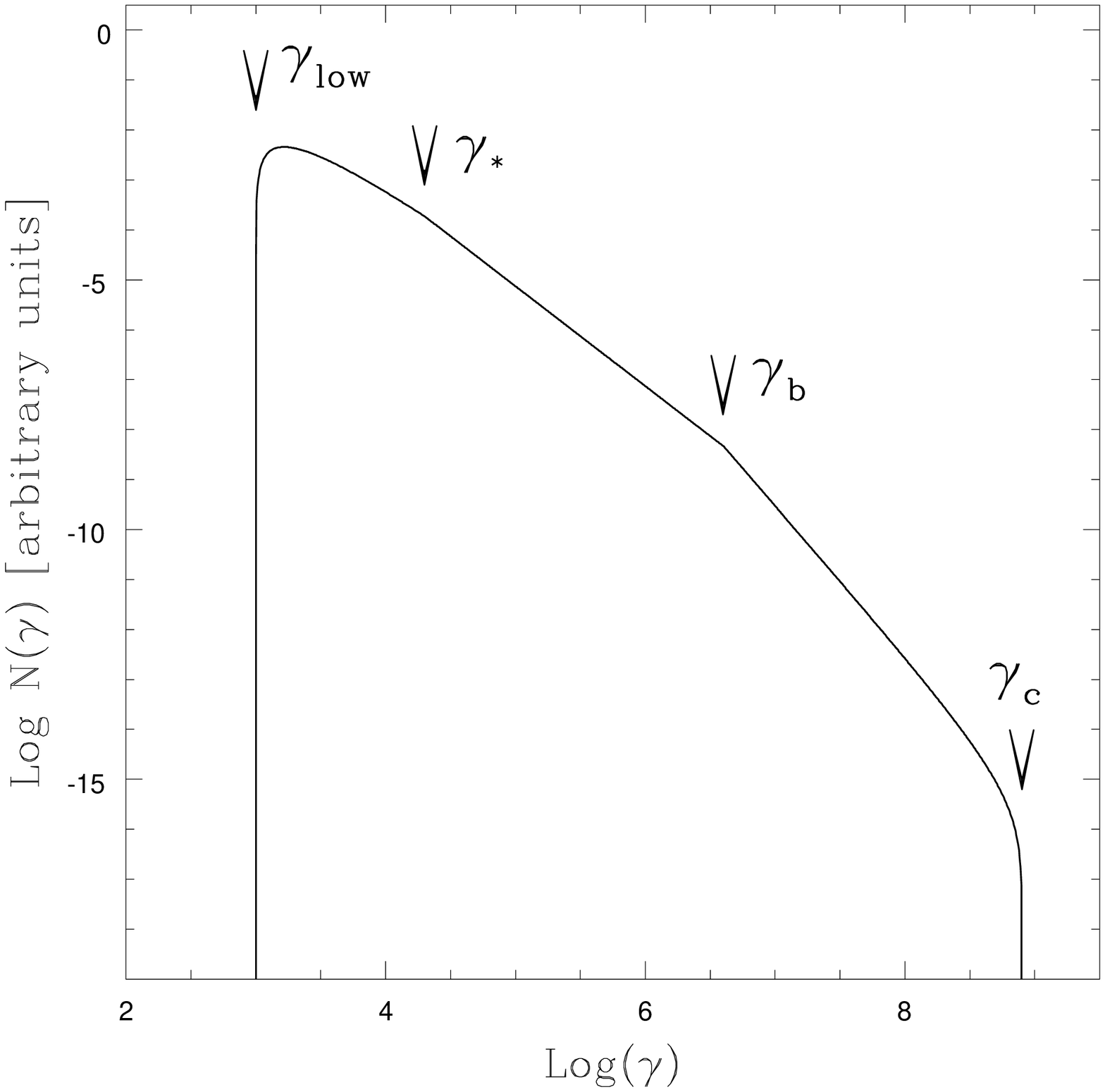}}
\caption{The electron spectrum integrated over the
post--shock region 
is reported as a function of the energy for 
a given set of parameters. 
The energies corresponding
to the relevant breaks
and cut--offs are indicated.}
\end{figure}

\subsection{Radiative processes and electron spectrum}

One of the basic motivations for this contribution 
is the possibility 
to study extragalactic radio sources
at different wavelength with comparable spatial
resolution which has come about in the last 2--3 years.
In this Section we will point out that 
broad band (possibly from radio to X--rays)
studies of the non--thermal emission
from radio sources can allow us to
derive unique information 
on the electron energy distribution. 

Based on theoretical argumentations, 
in the preceding Section we have shown that
the power law approximation for the
electron spectrum, 
usually adopted to calculate
the synchrotron and inverse Compton spectrum from 
extragalactic radio sources, is now inadequate.
In particular we have shown that cut--offs at low
and high energies, $\gamma_{\rm low}$ and 
$\gamma_{\rm c}$ respectively, 
might be expected if the relativistic electrons are accelerated 
in a shock region.
In addition, a break at intermediate energies, $\gamma_{\rm b}$,
is expected if the accelerated electrons age in a post--shock
region where acceleration processes are not efficient.
Finally, a low energy break, $\gamma_*$, or a flattening
of the spectrum, might be expected
depending on the energy distribution of the electrons 
(already relativistic) when
injected by the jet flow into the shock region.
The presence and the location of all the above 
breaks and cut--offs depend on the relative importance
of the mechanisms at work.
As a consequence, broad band studies of non--thermal emission
from the extragalactic radio sources, allow us 
not only to constrain the shape of the spectrum of the emitting
electrons, but they can also constrain the 
efficiency of the different mechanisms.

In what follows we concentrate on deriving
the energy of the electrons emitting in different frequency bands 
synchrotron and 
inverse Compton photons from compact regions (jets and hot spots)
and from extended regions (radio lobes).
For a general treatment of these emitting mechanisms, again, 
we refer the reader to
classical books (e.g., Ribicky \& Lightman 1979; Melrose 1980)
or to seminal papers (e.g., Blumenthal \& Gould 1970).

\subsubsection{Radio Lobes}

The most important radiative processes active in the radio
lobes are the synchrotron emission and the IC scattering
of external photons. As the radio lobes are generally regions
extended tens or hundreds of kpc,
the IC scattering of the synchrotron 
radiation produced by the same electron population (SSC) 
is not in general an efficient process.

{\it SYNCHROTRON EMISSION}: the typical energy of the electrons
responsible for synchrotron emission observed at
radio, optical and X--ray
frequencies is given by:

\begin{eqnarray}
\gamma_{\rm syn} \sim
{1\over 2} 
{{ 10^3 }\over{ {B_{\mu G}}^{1/2} }} {\nu_{\rm MHz}}^{1/2}
\sim 5000 \nonumber\\
\sim 1 \times 10^6 \nonumber\\
\sim 4 \times 10^7
\label{synene2}
\end{eqnarray}

where we have considered a magnetic field strength of the order
of $\sim 30 \mu$G, typical of the lobes of
relatively powerful FRII radio galaxies.

{\it IC EMISSION}: the relativistic electrons in the radio lobes can IC scatter
external photons to higher frequencies.
The nature of the external photons depends on the astrophysical
situation we are considering. 
Here we focus on the IC scattering
of the CMB photons and on the IC scattering of the 
nuclear (quasar, QSO) photons.

$\bullet$
The IC scattering of CMB photons into the X--ray band 
is a well known process (e.g., Harris \& Grindlay 1979)
and it has been successfully revealed in a few 
radio galaxies (Section 4.1). 

Optical emission from IC scattering
of CMB photons by lower energetics electrons 
($\gamma \sim 20$) was also tentatively 
proposed in order to account for 
the {\it optical-UV alignment} discovered 
in high--z radio galaxies (e.g., Daly 1992).
So far, to our knowledge, there are no cases
of detection of this effect. 
This may possibly be related to the  
expected flattening of the electron spectrum at 
$\gamma < 50$ due to the Coulomb losses (Sect.5).

$\bullet$
More recently, Brunetti et al. (1997) have proposed that the
IC scattering of nuclear photons into the X--ray
band might be an efficient process in powerful and relatively
compact FRII radio 
galaxies and quasars.
Powerful nuclei can isotropically 
emit up to $\sim 10^{47}$erg/s 
in the far-IR to optical band so that their
energy density typically outweight that due to the CMB 
for $\leq 100$ kpc distance from the nucleus.

The energy of the electrons emitting in the optical and X--ray 
band due to IC scattering is:

\begin{eqnarray}
\gamma_{\rm ic}
\sim \left( {{ \nu }\over{\nu_{\rm E}}} \right)^{1/2}
\sim 1000\,\,\,\,\,\,\,\,\,\,\,\,\,\,\,
( {\rm CMB} \rightarrow {\rm X-ray} )
\nonumber\\
\sim 20\,\,\,\,\,\,\,\,\,\,\,\,\,\,\,\,\,
( {\rm CMB} \rightarrow {\rm optical} )
\nonumber\\
\sim 100\,\,\,\,\,\,\,\,\,\,\,\,\,\,\,\,
( {\rm QSO} \rightarrow {\rm X-ray} )
\label{eicene2}
\end{eqnarray}

$\nu_{\rm E}$ and $\nu$ being the frequency of the external (seed) 
and of the scattered photons, respectively.

Both IC/CMB and IC/QSO can produce detectable
X--rays from the radio lobes which might introduce some degeneracy
in the interpretation of the observed fluxes.
However, it should be noticed that the contribution from the above
IC processes can be easily disentangled based on the morphology
of the observed X--ray emission.
Indeed, in the IC/QSO model the photons propagate from the nucleus
and their momenta are not isotropically distributed when
they scatter with the electrons in the lobes: i.e.,
the
IC scattering process is anisotropic.
It follows that at any given energy of the scattered
photons there will be many more scattering events when the
velocities of the relativistic
electrons point to the nucleus,
(i.e. the direction of the incoming photons) than
when they point in the opposite direction.
The resulting IC emission will be enhanced towards the
nucleus and will be
essentially absent in the opposite direction.
As a consequence, if the radio axis of the radio source
does not lie on the plane of the
sky, the X--rays from IC/QSO will be asymmetric:
the smaller the angle between the axis and the line of sight,
the greater the difference in IC emission from two
identical lobes (Brunetti et al. 1997; Brunetti 2000).

In Fig.5 we report a compilation of the energy ranges
selected by the different emitting processes in the different 
bands superimposed on representative electron spectra 
produced in different acceleration scenarios.
\begin{figure}
\resizebox{\hsize}{!}{\includegraphics{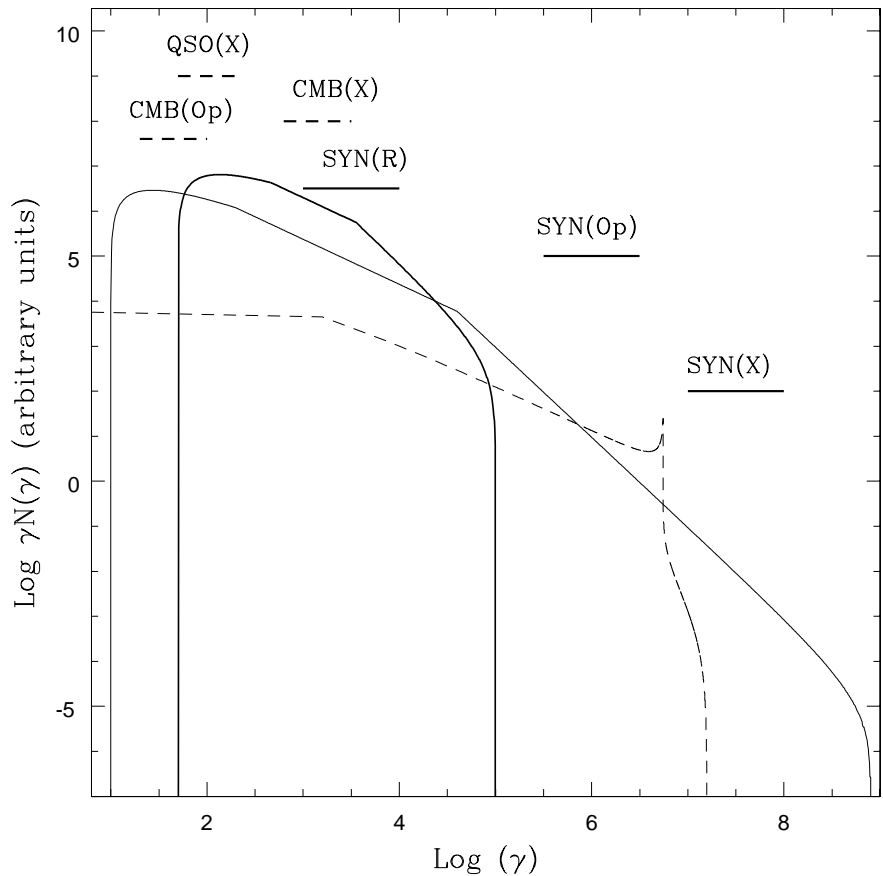}}
\caption{Scheme of the sampling of the spectrum of the
emitting electrons with the different processes 
(SYN= synchrotron, CMB= IC/CMB, QSO= IC/QSO)
in different bands (R= Radio,
Op= Optical, X= X--ray)
in the case of extended regions.
The different electron spectra are shown for a
set of different acceleration models.}
\end{figure}
As a net result, the combination of
studies of the spectra emitted by 
different processes at different frequencies, will allow 
us to derive information/constraints on the electron spectrum
which cover all the relevant energies.

\noindent
In particular, it should be stressed that :

$\bullet$
In general,
the detection of synchrotron radiation from the radio 
lobes at high frequencies (optical or X--rays)
requires $\gamma > 10^5$ emitting electrons.
Due to the radiative losses, these electrons have 
life times which are at least one order of magnitude 
shorther than the dynamical time/age of the 
radio lobes.
As a consequence, the detection of such emission 
would prove the presence of effective in situ
reacceleration (or injection) 
mechanisms distributed over whole the volume of 
the radio lobes.
To our knowledge, so far, there are no 
clear detection of synchrotron
optical (or even X--ray) emission 
from the lobes of radio galaxies and quasars.

$\bullet$
the X--ray emission from the IC/CMB is contributed by
electrons in an energy range not far from that
of the electrons emitting synchrotron radiation 
in the radio band ($\gamma \sim 10^3 - 10^4$).
As it will be pointed out in the next Section, this helps
constraining the number density of the relativistic electrons 
in the case where the spectrum of the relativistic 
electrons is not well known {\it a priori}.

$\bullet$
the X--ray emission produced by the IC/QSO 
is produced by low energy electrons 
($\gamma \sim 100$).
This is very important as the detection and 
study of the spectrum of this 
emission allows us to constrain the energy distribution of the
emitting electrons at very low, and still unexplored energies, 
which might contain most of the energetics of the radio lobes.

\subsubsection{Jets and hot spots}

Although radio hot spots are believed to advance 
in the surrounding medium 
at non--relativistic velocities 
(e.g., Arshakian \& Longair 2000),
there are several indications that radio jets are
moving at relativistic speeds out to several tens
of kpc distance from the nucleus (e.g., Bridle
\& Perley, 1984; see
also T. Venturi, these proceedings).
As a consequence, in this Section, we assume 
that the photons emitted by the jet 
are observed at a frequency ${\cal D} \nu_o$,
where $\nu_o$ is the emitted (i.e., jet frame), 
frequency and ${\cal D}$ the
Doppler factor, i.e. :

\begin{equation}
{\cal D}= { 1 \over{ \Gamma ( 1 - 
{{ v_{\rm jet} }\over {c}} \cos \theta_{\rm jet} ) }}
\label{doppler}
\end{equation}

where $\Gamma$ is the Lorentz factor of the
jet, $v_{\rm jet}$ is the velocity of the jet and
$\theta_{\rm jet}$ is the angle between the direction
of the velocity of the jet and the line of sight.

$\bullet$
In the case of a jet moving close to the direction
of the observer (we assume
${\cal D}\sim \Gamma$), the typical energy 
of the electrons 
responsible for synchrotron emission observed at 
X--ray, optical and radio frequencies is given by:

\begin{eqnarray}
\gamma_{\rm syn}\sim 
{{ 2 \cdot 10^8 }\over{\Gamma^{1/2}}}
\left( {{ \epsilon ({\rm keV}) }\over{B_{\mu G}}} \right)^{1/2}
\sim 3 \times 10^6 \left( {{ 10 }\over{\Gamma}} {{ 500 }\over{
B_{\mu G}}} \right)^{1/2} \nonumber\\
\sim 10^5 \left( {{ 10 }\over{\Gamma}} {{ 500 }\over{
B_{\mu G}}} \right)^{1/2} \nonumber\\
\sim 400 \left( {{ 10 }\over{\Gamma}} {{ 500 }\over{
B_{\mu G}}} \right)^{1/2}
\label{synene}
\end{eqnarray}

\noindent
In Fig.6 we report the detailed calculation of the
energies of the electrons 
responsible for the synchrotron emission observed in different
frequency bands as a function of the angle $\theta_{\rm jet}$.

\begin{figure}
\resizebox{\hsize}{!}{\includegraphics{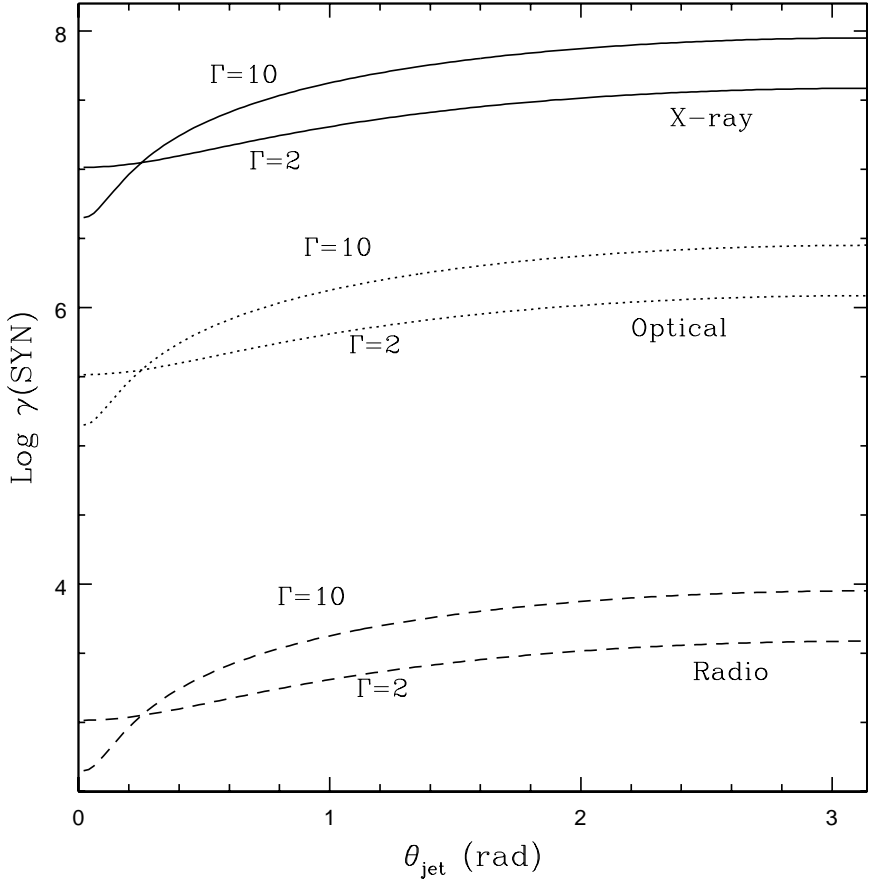}}
\caption{The energy of the electrons
responsible for the synchrotron emission observed at radio
(solid lines), optical (dotted lines) and X--ray
(dashed lines) frequencies is reported 
as a function of the angle $\theta_{\rm jet}$ and
for relevant values of the bulk Lorentz factor
$\Gamma$. The plots are calculated assuming
$B=100 \mu$G in the jet.}
\end{figure}

$\bullet$
A relevant radiative process in the case of 
compact regions is the inverse Compton scattering
of the synchrotron photons emitted by the same electron
population (SSC, e.g., Jones et al. 1974a,b;
Gould 1979).
In this case, the energy of the electrons giving
the SSC emission observed at X--ray and optical
frequencies is:

\begin{eqnarray}
\gamma_{\rm ssc}\sim
 \left( {{\nu}\over{\nu_{\rm R}}}
\right)^{1/2} \sim 1.5 \times 10^4
\nonumber\\
\sim 500
\label{sscene}
\end{eqnarray}

which 
does not depend on the motion of the jet with respect
to the observer. This is because the frequency of
both the
observed (seed) synchrotron photons $\nu_{\rm R}$,
and the observed
(scattered) inverse Compton photons $\nu$, scale with
the Doppler factor of the jet.

$\bullet$
An additional radiative process is the inverse
Compton scattering of external photons (e.g., Dermer 1995).
In the case of a jet moving approximately in the direction
of the observer, the energy of the electrons responsible
for the inverse Compton emission observed at optical and
X--ray frequencies is:

\begin{eqnarray}
\gamma_{\rm eic} \sim
{1\over{\Gamma}} \left( {{ \nu }\over{\nu_{\rm E}}}
\right)^{1/2} \sim 100 \times \left( {{ 10 }\over{\Gamma}}
\right) \left( {{ \nu_{\rm cmb}}\over{\nu_{\rm e}}}
\right)^{1/2} \nonumber\\
\sim 3 \times \left( {{ 10 }\over{\Gamma}}
\right) \left( {{ \nu_{\rm cmb}}\over{\nu_{\rm e}}}
\right)^{1/2}
\label{eicene}
\end{eqnarray}

\noindent
where we have parameterized the frequency of
the external photons $\nu_{\rm e}$, with that
of the CMB photons $\nu_{\rm cmb}$, whose energy density
becomes relevant if the
jet is moving at high relativistic speed.

\begin{figure}
\resizebox{\hsize}{!}{\includegraphics{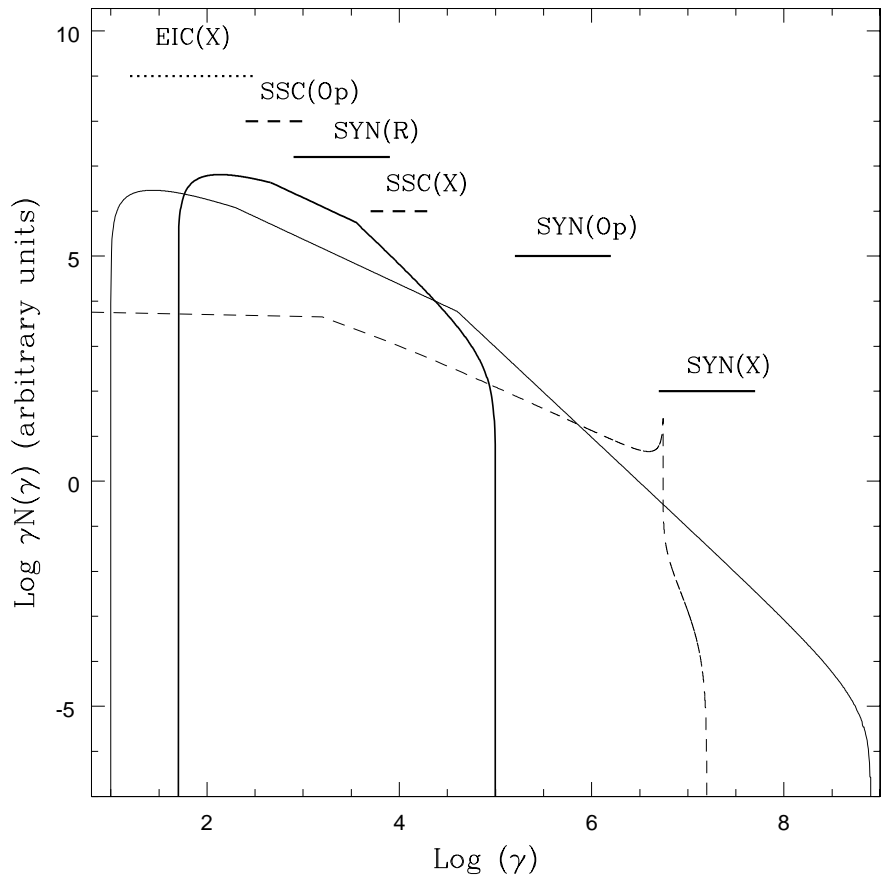}}
\caption{Scheme of the sampling of the spectrum of the
emitting electrons with the different processes
(SYN= synchrotron, EIC= IC/CMB assuming $\Gamma=10$ and
a jet moving in the direction of the observer, SSC= 
Syncro--self--Compton)
in different bands (R= Radio,
Op= Optical, X= X--ray)
in the case of compact regions.
The different electron spectra are shown for a
set of different possible acceleration models.}
\end{figure}

In Fig.7 we report a compilation of the energy ranges 
selected by the different emitting processes in the different bands
compared with representative electron spectra produced in different
acceleration scenarios.
As found in the case of the extended regions, studies of the
spectra emitted by the different processes at 
different frequencies allow us to
derive information/constraints on the electron spectrum.

\noindent
In particular, it should be noticed that :

$\bullet$
The detection of synchrotron radiation from large scale jets 
at X--ray frequencies requires $\gamma > 10^6$ 
emitting electrons which, due to the radiative
losses, have a relatively short life time and thus their 
presence points to the need for reacceleration 
(or injection) mechanisms 
in these jets.
This requirement is even stronger in the case of high frequency
synchrotron emission from jets moving at small angles 
with respect to the plane of the sky (e.g., in case of
narrow line radio galaxies or FR I with quasi--symmetric jets). 
In the case of the hot spots, which are believed to advance
at non relativistic speeds, the detection of
synchrotron radiation in the optical band is very important as
it can be considered as a prove for the presence of efficient
in situ reacceleration mechanisms in these regions.

$\bullet$
If the emitting region is roughly homogeneous (in $B$ field), 
the X--ray emission due to SSC process is produced 
by roughly the same electrons ($\gamma \sim 10^3 - 10^4$) 
emitting synchrotron radiation in the radio band.
As it will be pointed out in the next Section, this would help
in constraining the number density of the relativistic electrons 
in the case in which the energy
distribution of the relativistic electrons is unknown
{\it a priori}.

$\bullet$
As the IC scattering of the external (e.g., CMB)
photons is expected to be 
efficient only if the jets is moving in the direction of the 
observer at relativistic speed (e.g., Harris \& Krawczynski 2002), 
the relative X--ray emission should be produced by low 
energy electrons ($\gamma \sim 100$).
This is very important as the spectrum of the observed X--rays 
provides constraints to the energy distribution of the
emitting electrons at very low and unexplored energies.
This is similar to the case of the IC/QSO from the radio lobes.

\subsection{Methods to constrain the energetics of 
extragalactic radio sources}

One of the basic goals of radio 
astrophysics is to 
constrain the energetics of the extragalactic radio sources
and the ratio between the energy density of particles and
fields. 
Assuming for simplicity an electron energy distribution 
$N(\gamma)=K_e \gamma^{-\delta}$, 
the energy density of the relativistic plasma is :

\begin{equation}
\omega (e+p+B)=
{\rm m} {\rm c}^2 K_e (1+k) \int_{\gamma_{\rm low}}^
{\gamma_{\rm max}} d\gamma \gamma^{-\delta + 1}
+ {{ B^2}\over{8 \pi}}
\label{energy}
\end{equation}

\noindent
where $k$ is the ratio between the energy of protons
and electrons.

In what follows we describe the two most usually adopted
arguments to constrain the energy density of the emitting 
plasma: the minimum energy assumption and the IC method.
We refer the reader to classical books
(e.g., Verschur \& Kellermann, 1988) for additional
methods (e.g., synchrotron self absorption).

\subsubsection{Minimum energy assumption}

As it is well known, the synchrotron 
properties result from a complicated convolution
of magnetic field intensity and geometry with the electron
spectrum.
As a consequence, radio synchrotron observations alone are generally
insufficient to derive the relevant physical parameters
(e.g., Eilek 1996; Eilek \& Arendt 1996;
Katz--Stone \& Rudnick 1997, Katz--Stone et al. 1999); 
this problem
is known as the {\it synchrotron
degeneracy}.

Hence, 
to have an idea of the energetics of
extragalactic radio sources, 
radio astronomers are forced to
calculate the minimum energy
conditions: i.e., the minimum energy 
of the relativistic plasma required to
match the observed synchrotron properties.
For the details of this classical argument 
we refer the reader to Packolzyick
(1970) and Miley (1980).
Here, we will
briefly focus on a critical review of the minimum
energy formulae as usually adopted.
The magnetic field strength yielding the minimum energy
(equipartition field) is:

\begin{equation}
B_{\rm eq} = C_{\rm Pa}(\alpha)
(1+k)^{2/7} 
\left( {{\int_{\nu_1}^{\nu_2}L_{\rm syn}(\nu)d\nu }\over{ V }}
\right)^{2/7}
\label{beq1}
\end{equation}

\noindent
where the constant $C_{\rm Pa}$ can be derived from Packolzyik
(1970). Eq.(\ref{beq1}) is widely used by radio astronomers 
to calculate the minimum energy conditions
in a radio source.
For historical reasons the frequency band 
usually adopted to
calculate the equipartition field is $\nu_1=$10 MHz -- $\nu_2=$100 GHz, 
i.e.
roughly the frequency range observable with 
the radio telescopes.
From a physical point of view, the adoption of this
frequency band means that, in the 
calculation of the minimum energy, 
it is assumed that only electrons emitting between 10 MHz --
100 GHz, i.e. with energy between
$\gamma_{\rm low}= a (\nu_1/B_{\rm eq})^{1/2}$
and $\gamma_{\rm max}= a (\nu_2/B_{\rm eq})^{1/2}$, 
are present in the radio source.

\noindent
This generates a physical bias because :

i) There are no physical reasons which point to the absence of
electrons at lower energies and 
a substantial fraction of the energetics might
be associated to $\gamma < 1000$ electrons which usually
emit synchrotron radiation below 10 MHz.

ii) As the energy of the electrons which emit synchrotron radiation
at a given frequency depends on the magnetic field intensity,
the low frequency cut--off, $\nu_1$, in Eq.(\ref{beq1})
corresponds to a low energy cut--off, $\gamma_{\rm low}$, 
in Eq.(\ref{energy}) which depends on the magnetic field
strength $B_{\rm eq}$. As a net result,
in radio sources with different $B_{\rm eq}$, 
the classical minimum
energy formulae select different energy bands of the
electron population.

Although point i) can be recovered introducing
a very low frequency cut--off ($\sim 1-100$ kHz) in 
Eq.(\ref{beq1}),
it is not possible to recover the ii) with the classical
equations (e.g., Myers \& Spangler, 1985;
Leahy 1991).
Based on these considerations, in this contribution, 
we follow a different approach to calculate the minimum energy 
conditions which 
does not use a fixed emitted frequency band.

Expressing $K_e$ in Eq.(\ref{energy}) in terms of the 
monochromatic synchrotron luminosity, $L_{\rm syn}(\nu)$,
and of the magnetic field intensity, Eq.(\ref{energy}) can 
be immediately minimized as a function of the magnetic field.
For $\alpha>0.5$, 
the value of the magnetic field yielding the minimum
energetics is (Brunetti et al., 1997):

\begin{equation}
B_{\rm eq} = \left[ C(\alpha)  (1+k)
{{ L_{\rm syn}(\nu) \nu^{ \alpha } }\over{
V }} \right]^{ {1\over{\alpha +3}}}
{\gamma_{\rm min}}^{ {{ 1-2\alpha}\over{\alpha + 3}}}
\label{beq2}
\end{equation}

which does not depend on the emitted frequency
band but directly on the low energy cut--off of
the electron spectrum. 
The ratio between the equipartition
magnetic field, $B_{\rm eq}$ (Eq.\ref{beq2}), 
and the classical
equipartition field, $B^{\prime}_{\rm eq}$
(Eq.\ref{beq1}), is given by:

\begin{equation}
{{ B_{\rm eq}}\over{B^{\prime}_{\rm eq}}}
\sim 1.2 \left( B^{\prime}_{\rm eq}
\right)^{ {{ 1-2\alpha}\over{2(\alpha+3)}}}
{\gamma_{\rm min}}^{ {{ 1-2\alpha}\over{\alpha + 3}}}
\label{ratioeq}
\end{equation}

\noindent
As Eq.(\ref{beq2}) selects also 
the contribution to the energetics
due to the low energy electrons, the 
deriving intensity of the 
equipartition field is greater than that
of the classical field. 
In addition, such a difference increases with decreasing
magnetic field intensity of the radio sources and
in the case of steep electron energy distributions (Fig.8).

\noindent
Finally,
it can be shown that, the ratio between particle and field
energy densities is:

\begin{equation}
{{ \omega_{\rm eq}(e+p)}\over
{\omega_{\rm eq}(B)}}=
{ 2\over{\alpha+1}}
\label{ratioed}
\end{equation}

\noindent
which is $4/3$, i.e. the constant
ratio obtained with the
classical formulae, in the case $\alpha =0.5$.

\begin{figure}
\resizebox{\hsize}{!}{\includegraphics{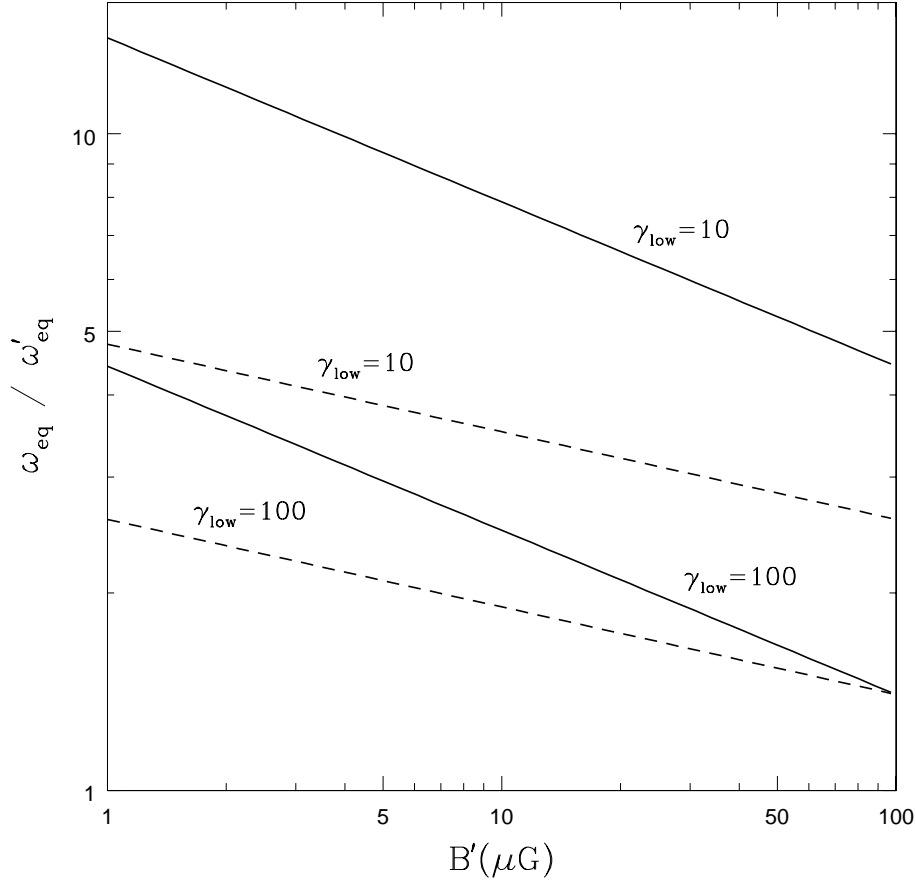}}
\caption{The ratio between the energy density 
of the equipartition field (Eq.\ref{beq2})
and of the classical equipartition field
(Eq.\ref{beq1}) is shown.
The ratio is reported as a function of the strength of 
the classical field $B^{\prime}_{\rm eq}$,
for different values of $\gamma_{\rm low}$
and of $\delta$=2.5 (dashed lines) and 3.0 (solid lines).
}
\end{figure}

It should be reminded that 
in Section 4.2 and 4.3 we have shown that the electron 
energy distribution is not expected to be a simple power 
law and thus Eq.(\ref{beq2}) is in general only an 
approximate solution.
In order to calculate the equipartition field
for a general electron energy distribution a numerical
approach is required.

\subsubsection{Inverse Compton method}

Already many years ago it was pointed out
(e.g., Felten \& Morrison 1966) that 
the {\it 
synchrotron degeneracy} in determining the physical
properties of radio sources could be broken by measurements
of the X--rays produced by inverse Compton scattering.

It is well known (e.g., Blumenthal \& Gould 1970) that 
the IC emissivity depends on the number and spectrum of
the scattering electrons and of the incident photons. 
For a power law energy distribution
of the electrons, one has:

\begin{equation}
L_{\rm ic}(\nu_{\rm ic})=
K_e V C_{\rm ic}(\alpha,n(\nu_{\rm ph}),..)
\nu_{\rm ic}^{-\alpha}
\label{lic}
\end{equation}

$C_{\rm ic}$ can be derived in a number
of general cases (isotropica case:  
e.g., Blumenthal \& Gould 1970, anisotropic case: 
e.g., Aharonian \& Atoyan 1981; Brunetti 2000).
Given the energy (and angular)
distribution of the incident photons $n(\nu_{\rm ph})$, 
the measure of the IC flux and spectrum allows to 
constrain the electron number density and energy 
distribution.

For a power law energy distribution the synchrotron
luminosity is given by (e.g., Ribickyi \& Lightman 1979):

\begin{equation}
L_{\rm syn}(\nu_{\rm syn})=
K_e V C_{\rm syn}(\alpha)B^{\alpha+1}
\nu_{\rm syn}^{-\alpha}
\label{lsyn}
\end{equation}

so that combining Eqs.(\ref{lic}) and (\ref{lsyn}) one
can derive the intensity of the magnetic
field :

\begin{equation}
B= \left[
{{ L_{\rm syn} }\over{ L_{\rm syn} }}
{{ C_{\rm ic} }\over{ C_{\rm syn} }}
\right]^{1/(\alpha+1)}
\left( {{ \nu_{\rm syn} }\over{ \nu_{\rm ic} }}
\right)^{\alpha/(\alpha+1)}
\label{bic}
\end{equation}

\noindent
Once $B$ is estimated, 
the ratio between particle and field energy density
in the emitting region is given by :

\begin{equation}
{{ \omega(e+p) }\over
{\omega(B)}}=
{2\over{\alpha +1}}
\Delta^{\alpha +3}
\label{ratiobic}
\end{equation}

where $\Delta = B_{\rm eq}/B$.
Under equipartition conditions (i.e. $\Delta=1$),
Eq.(\ref{ratiobic}) is equivalent 
to Eq.(\ref{ratioed}).

If the assumption of a power law energy distribution is relaxed,
Eq.(\ref{bic}) should be replaced with a more complicated
formula. Nevertheless, an approach to 
the determination of the magnetic field intensity is
always possible.
The most famous applications of this method  
are those of the 
IC scattering of CMB photons (e.g., Harris \& Grindlay 1979)
whose energy density and spectrum $n(\nu_{\rm ph})$,
are well known, and of the SSC emission from
compact regions (e.g., Jones et al. 1974a,b; Gould 1979).
In these cases, as noticed in the previous Section, the
radio synchrotron and the X--ray IC photons are 
emitted by about the same electrons and the value
of the $B$ field poorly depends on the electron spectrum
(i.e., Eq.(\ref{bic}) is always applicable).
A new application of the IC method has
been proposed in the case of IC scattering of
nuclear photons from the radio lobes (Brunetti et al.,
1997).

\section{THE NEW OBSERVATIONS}

In this Section we will try to give a possibly `unbiased' 
review of some of the most recent observations which are helping 
us to better understand the physics of extragalactic radio sources.
Especially in the first part of this review, where we focus
on recent detections of IC scattering from the lobes
of radio galaxies and quasars, the number of observations is
still very small. 
Consequently, unbiased considerations on the physics of radio
lobes will be only obtained in the future thanks to the
expected improvement of the statistics.

\subsection{X--ray observations of IC/CMB emission 
from radio lobes}

Despite the poor spatial resolution and sensitivity,
non--thermal IC/CMB
X--ray emission from the radio lobes has been discovered 
by {\tt ROSAT} and {\tt ASCA} in a few nearby radio galaxies, 
namely Fornax A (Feigelson et al. 1995;
Kaneda et al. 1995; Tashiro et al. 2001),
Cen B (Tashiro et al. 1998),  
3C 219 (Brunetti et al. 1999) and NGC 612 (Tashiro et al.,
2000).
By combining X--ray, as IC scattering of CMB photons,
and synchrotron radio flux densities
it was possible to derive magnetic field strengths
(averaged over the total radio volume) 
0.3--1 times the equipartition fields.
In general, these observations have been complicated due to
the weak X--ray brightness, relatively low count statistics and
insufficient angular resolution of the instruments.

After approximately three years of observations with {\it Chandra}
and XMM--{\it Newton} no clear evidence for diffuse emission
from IC scattering of CMB photons from the lobes of 
extragalactic radio sources has been published on a referred journal.
However, future to our knowledge there are preliminary 
results with {\it Chandra}
and with XMM--{\it Newton} which appear very promising and to which
the reader is referred at the epoch of the publication of this
contribution.

\subsubsection{3C 219}

So far, the only public (on electronic preprint archive)
case of IC/CMB detection 
is 3C 219 (Brunetti et al.2002a, astro-ph/0202373).
3C 219 is a nearby (z=0.1744) powerful FRII radio galaxy 
extending for $\sim$ 180 arcsec corresponding to a
projected total size of $\sim$ 690 kpc.
The {\it Chandra} (17 ksec)
0.3--8 keV image is shown in Fig.9a superimposed
on the radio contours from a deep 1.4 GHz VLA observation.  
Thanks to the arcsec resolution, it was possible to 
disentangle the nuclear emission (which affects only the 
innermost 3--5 arcsec) from the other components, 
and to identify 
the bright clump visible 
on the north--west with a background cluster at z=0.39.
\begin{figure}
\resizebox{\hsize}{!}{\includegraphics{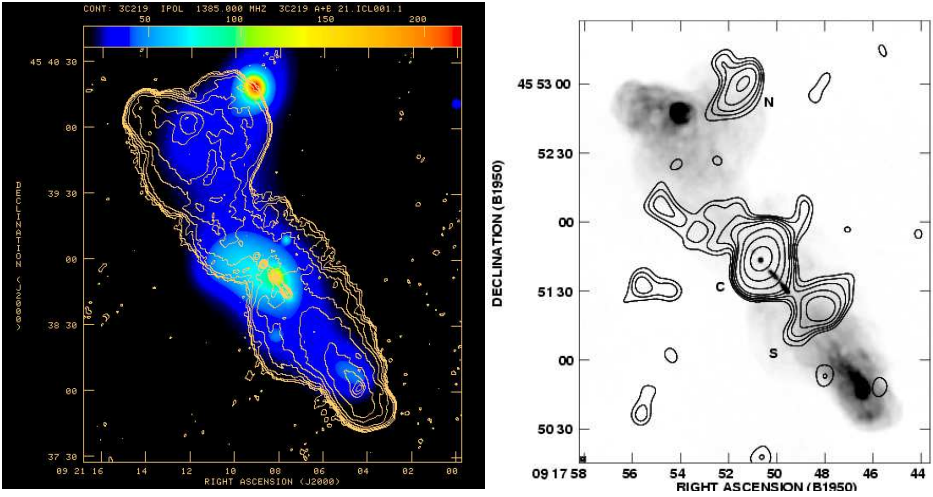}}
\caption{
{\bf Panel a}:
Radio VLA contours at 1.4 GHz of 3C 219 superimposed
on the {\it Chandra} 0.3--8 keV image ({\bf grays}). 
The X--ray image is smoothed after removal of the nuclear 
source which affects only the innermost 3-5 arcsec.
{\bf Panel b}:
X--ray {\tt ROSAT HRI} image ({\bf contours}) obtained
after the subtraction of the nuclear source is overlaid with
a VLA radio image ({\bf grays}). The subtraction of
the X--ray nucleus strongly affects the X--ray contours
in the innermost 20 arcsec.}
\end{figure}
Diffuse emission coincident with the radio lobes
also showing a brightness increment in the innermost part of the
northern lobe is clearly detected.

The combined imaging and spectral analysis of this emission
($\sim 400$ net counts) point to a non--thermal, IC/CMB origin
of the large scale diffuse emission.
Following the procedure described in Section 3.5.2, 
the comparison between radio and X--ray IC emission yields a precise
measurement of the magnetic field intensity (averaged over the total
radio volume) which results a factor of $\simeq 2.9$ times lower than 
the equipartition value (assuming $\gamma_{\rm low}=50$ in 
Eq.(\ref{beq2}).
Under these conditions, the ratio between particle and field
energy densities (Sect.3.5.2) is $\simeq 64$.
The derived energetics of the lobes of 3C 219 is
$\simeq 10^{60}$erg which results 
a factor $\simeq 7$ larger than that estimated 
with classical equipartition formulae (Sect.3.5.1).
The bulk of the energy density of the radio lobes
is associated to the electrons with $\gamma < 500$.

The increment in the X--ray brightness present
in the innermost part 
($\sim 70$ kpc) of the northern lobe (counter lobe)
may indicate an additional contribution due to  
IC scattering of nuclear photons,
thus providing direct evidence for the presence of $\gamma \sim
10^2$ electrons in the lobes.
Finally, 
two distinct knots at 10--25 arcsec south of the nucleus, 
spatially coincident with
the radio knots of the main jet, are visible in the {\it Chandra}
image.

Past combined {\tt ROSAT PSPC}, {\tt HRI} 
and {\tt ASCA} observations did also find evidence for IC emission
in 3C 219 lobes out of equipartition conditions
(Brunetti et al. 1999).
However, the presence of the strong nuclear source, 
the impossibility to perform spatially resolved spectroscopy and 
the relatively poor sensitivity of {\tt ROSAT HRI}
required a follow up observation with {\it Chandra}.
The 0.1--2 keV image from the 30 ksec {\tt ROSAT HRI} observation
is shown in Fig.9b: the emission within  
$\sim 20$ arcsec from the nucleus is strongly affected
by the subtraction of the nuclear source.
The comparison of the two images in Fig.9 is very instructive
and shows well the real breakthrough in X--ray imaging  
provided by {\it Chandra}.

\subsection{X--ray observation of IC scattering of nuclear 
photons from radio lobes}

The detection of X--ray 
emission from IC/QSO has been recently achieved 
in at least three objects (3C 179, 3C 207, 3C 295)
whereas possible evidences have been
suggested in other few cases 
(e.g., 3C 294: Fabian et al. 2001; 3C 219, Fig.9).
A positive detection of this effect with {\it Chandra} was a 
specific prediction of this model in the case that 
a substantial fraction of the energetics of the radio lobes
is associated to the low energy end of the electron spectrum.
X--ray emission from IC scattering of nuclear photons with the
relativistic electrons in the radio lobes is expected to be
particularly efficient in the case of relatively compact 
(i.e., $\leq 100$ kpc) and strong FRII radio galaxies and 
steep spectrum radio quasars.
This is due to the dilution of the nuclear flux with distance 
from the nucleus.

\subsubsection{Radio galaxies: 3C 295}

This is a classical FRII 
at the center of a rich cluster (z=0.461).
The X-ray data obtained with previous instruments 
({\it Einstein} Observatory: Henry \& Henriksen 
1986; ASCA: Mushotzky \& Scharf 1997;  ROSAT: 
Neumann 1999) only allowed the study of
the cluster emission.

3C 295 was the first FRII source observed by {\it Chandra}.
Harris et al. (2000) obtained the 0.3--7 keV image of this
radio galaxy in which the hot spots 
and nuclear emission were well separated from the surrounding 
cluster contribution.
The presence of
possible diffuse X--ray emission related to 
the radio lobes was first addressed by these authors.
However, the morphology and intensity of this emission 
resulted particularly uncertain
due to the presence of the bright nuclear source and of the
northern hot spot at $\sim 2$ arcsec distance.
Stimulated by the results of Harris et al. (2000), 
Brunetti et al. (2001b) performed a more detailed analysis
in order to maximize the information on the X--rays from
the radio lobes.

\noindent
In particular, these authors performed the spectrum of the nuclear
source which cames out to be highly absorbed by a column density
of $\sim 10^{23}$cm$^{-2}$ and thus almost absent in the
0.2--2 keV image.
This image is shown in Fig.10:
\begin{figure}
\resizebox{\hsize}{!}{\includegraphics{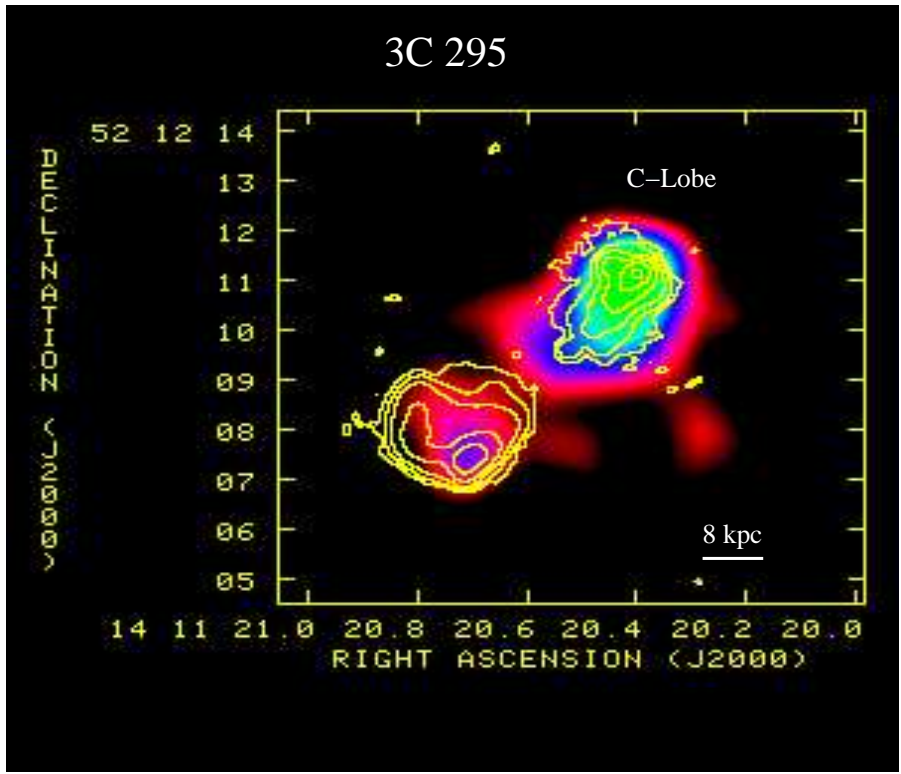}}
\caption{The GHz VLA radio map (contours) of 3C 295
is shown superimposed on the 0.1--2 keV 
{\it Chandra} image (grays).
The X--ray image is obtained
after the subtraction of the cluster emission.
}
\end{figure}
the morphology of the diffuse X--ray emission  
is double lobed with the X--rays coincident with the radio
lobes, thus pointing to a non--thermal origin.
In addition the asymmetry in the X--ray brightness (with the northern
lobe a factor $\sim 2-4$ brighter than the southern one)
appears to be the signature of the IC/QSO model.
In order to reproduce the observed brightness 
ratio with this model
an angle between radio axis and the plane of the sky of 6--13$^o$ is 
required with the northern lobe being further away from us.
This geometry was confirmed by the discovery of a faint 
radio jet in the southern lobe (i.e., the near one)
by P.Leahy with a deep MERLIN observation (private
communication).

As stated in Sect.3, the spectrum from 
IC scattering of nuclear photons is a unique tool to constrain the
energy distribution of $\gamma \sim 100-300$ electrons.
\begin{figure}
\resizebox{\hsize}{!}{\includegraphics{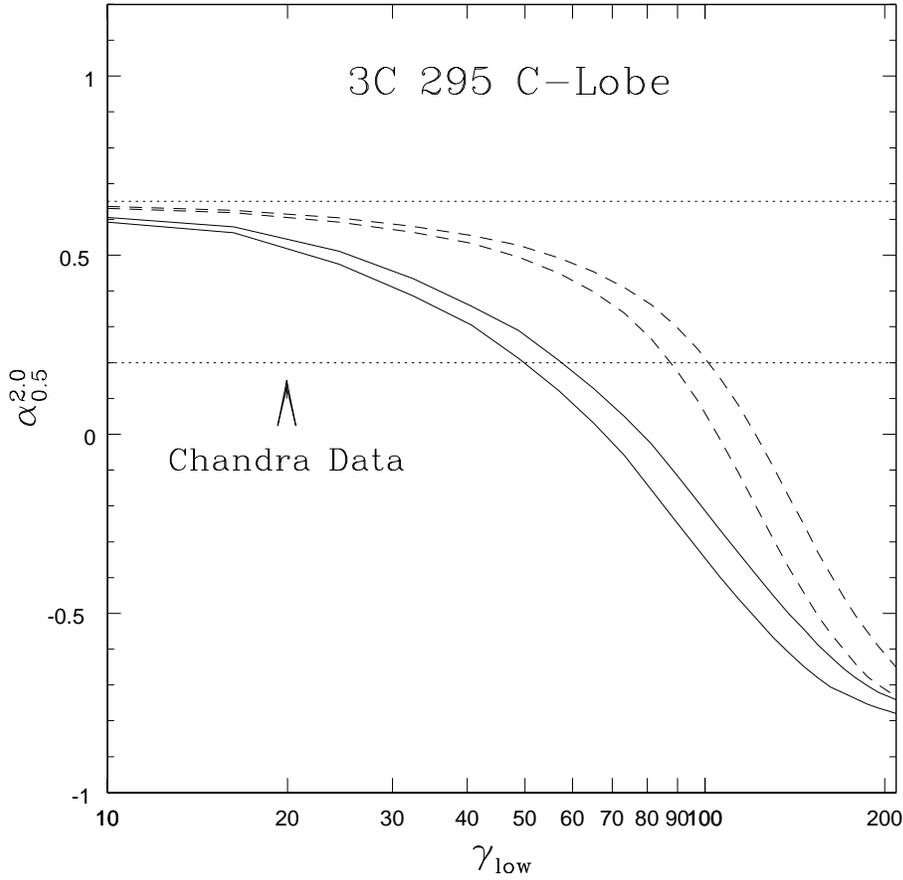}}
\caption{The 0.5--2 keV spectral index
expected in the case of the IC scattering of nuclear photons
is reported in the case of 3C 295.
In the calculation 
we have assumed the Sanders et al. (1989)
(solid lines) average SED of quasars and
the SED of 3C 48 (dashed lines).
For each SED we have reported two different curves
giving uncertainties on the spectral index
due to different set of parameters assumed 
in the model calculation.
The limits on the observed 0.5--2 keV spectral
index from our {\it Chandra} data
analysis are also reported (dotted lines).}
\end{figure}
Fig.11 shows the 0.5--2 keV spectral index predicted by
the model as a function
of a low energy cut--off in the electron spectrum : 
an upper limit $\gamma_{\rm low} <  100$ is obtained
from the {\it Chandra} data.

The IC scattering of the nuclear photons has been also
used to calculate the magnetic field strength in 
the lobes of 3C 295.
ISO measurement of 3C 295 flux (Meisenheimer et al. 2001) 
fix the far--IR nuclear luminosity to   
$\sim 3 \times 10^{46}$erg s$^{-1}$ and the deriving 
value of the IC magnetic field strength results 
consistent with the value calculated 
under minimum energy assumption (Fig.12).
The derived energetics of the radio lobes of 3C 295
is $\simeq 4.4\times 10^{59}$erg and results 
a factor $\sim 2$ larger than that derived with
classical equipartition formulae (Sect.3.5.1);
a significant fraction of it 
is associated to the $\gamma \leq 300$ 
electrons.

\begin{figure}
\resizebox{\hsize}{!}{\includegraphics{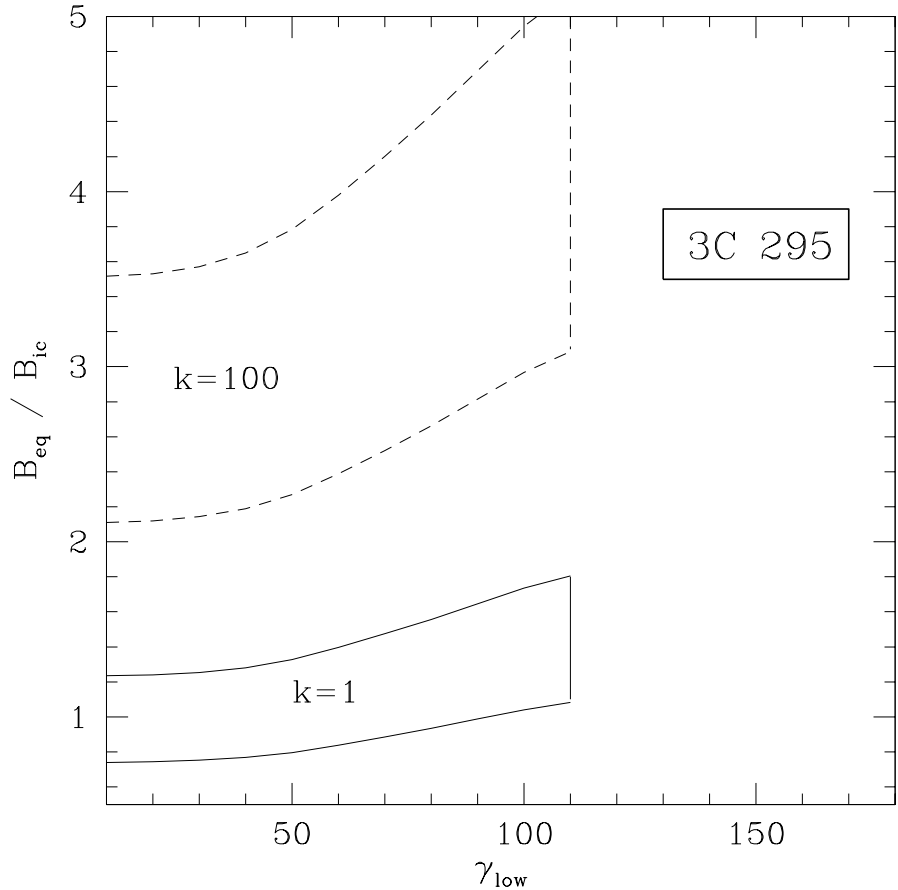}}
\caption{The ratio between the equipartition magnetic
field intensity ($B_{\rm eq}$) and that estimated
from the IC scattering is reported as a function
of the low energy cut--off $\gamma_{\rm low}$.
The calculation are performed for a bolometric
far--IR/optical luminosity of the hidden quasar
= 2--4$\times 10^{46}$erg s$^{-1}$;
k is the ratio between proton and electron energy density 
in the radio lobes.}
\end{figure}

\subsubsection{Lobe dominated quasars: 3C 179 and 3C 207}

The effect of the asymmetry in the X--ray distribution
from the anisotropic IC scattering
of the nuclear photons is maximized in the case of the
steep spectrum quasars, which typically 
make an angle of 10--30 degrees
between the radio axis and the line of sight.
This provides an unambiguous identification of the process responsible
for the X--ray emission as only the X--rays from the counter lobe
are expected to be efficiently amplified and thus detected.
In addition, in this case the far--IR to optical
flux from the nuclear photons 
can be directly measured thus allowing a prompt estimate of the
energy density of the scattering electrons (and magnetic field)
in the radio lobes as in the case of the IC scattering of CMB photons
(Sect.3.5.2).

So far there are two radio loud quasars observed with {\it Chandra}
in which extended X--ray emission from the counter--lobe has been
successfully detected, and for which no diffuse emission
from the near lobe was detected: 
3C 179 (Sambruna et al. 2002) and 
3C 207 (Brunetti et al. 2002b).
Both these sources are relatively compact and 
luminous, with prominent radio lobes making
them ideal candidates to detect IC scattering 
of the nuclear photons in
the radio lobes.

The 0.2--8 keV images of 3C 207 ($\sim$ 36 ksec exposure)
is reported in Fig.13 
superimposed on the VLA radio contours.
\begin{figure}
\resizebox{\hsize}{!}{\includegraphics{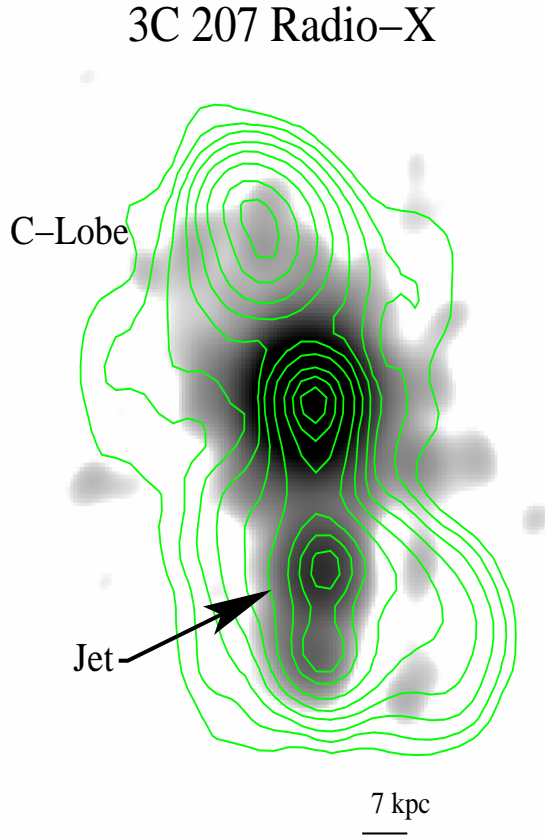}}
\caption{{\it Chandra} 0.2--8 keV image of 3C 207
(grays) superimposed on the 1.4 GHz VLA contours.
The X--ray jet and counter lobe are indicated in the
figure. The scale bar gives the resolution of the
X--ray image (0.9 arcsec = 7 kpc);
the resolution of the radio image is 1.4 arcsec.}
\end{figure}
\begin{figure}
\resizebox{\hsize}{!}{\includegraphics{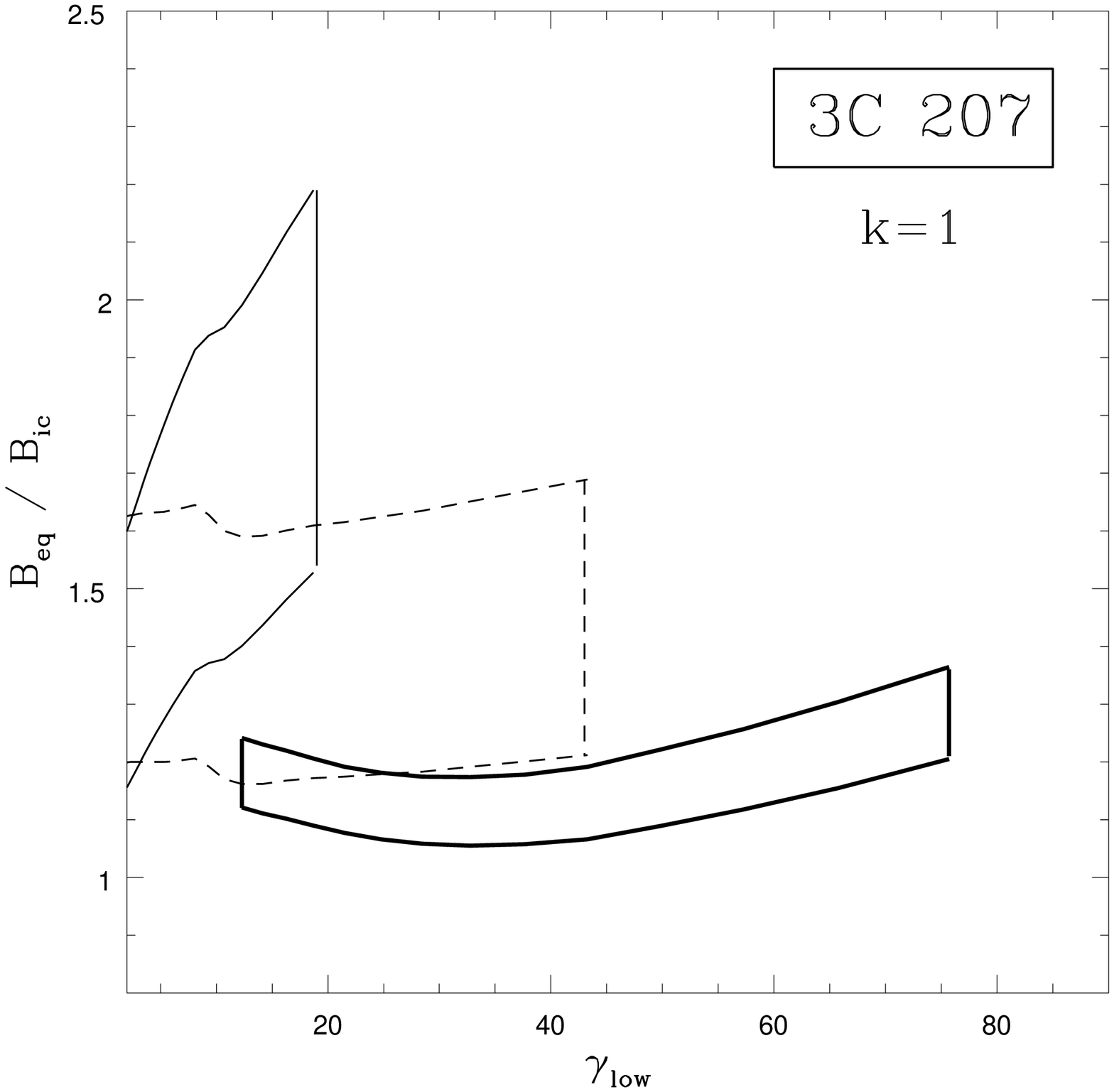}}
\caption{The allowance regions for the ratio between
IC--magnetic field and equipartition field strength
and for $\gamma_{\rm low}$ in the case of 3C 207
are reported for different assumed
energy distributions of the electrons:
power law $\delta=2.8$ down to $\gamma_{\rm low}$ (thick solid
region), accelerated spectrum with $\gamma_*/\gamma_{\rm low}$=2 
(dashed region) and =50 (thin solid region).
In the case of reaccelerated electron spectra, the
regions are calculated by assuming $\gamma_{\rm b}$ in the range
500--1000 and $\gamma_c >> 1000$.}
\end{figure}
The allowed regions of the values of the
magnetic field strengths
and of $\gamma_{\rm low}$ (Sect. 3.3)
as inferred by
the combined radio and X--ray fluxes and spectrum of 3C 207
are reported in
Fig.14: the magnetic field
strengths are lower, but within a factor of $\sim 2$,
from the equipartition values.
The resulting energetics of the radio lobes of 3C 207 
is $\simeq 3 \times 10^{60}$erg; a large fraction of it 
is associated to the $\gamma \leq 300$ electrons.
The above value of the
energetics is a factor $\sim 4.5$ larger than 
that obtained for 3C 207 with classical minimum energy 
formulae (Sect. 3.5.1).
The shorter {\it Chandra} exposure ($\sim$ 9 ksec) in the case
of 3C 179 makes difficult to constrain the energetics of the radio 
lobes. However, also in this case, the detection of IC/QSO
proves that a consistent fraction of the energetics of the
electrons component in the radio lobes is associated 
to low $\gamma$ electrons.

\subsection{Jets and hot spots}

{\it RADIO OBSERVATIONS}:
Radio telescopes have imaged a large number of jets and
hot spots of radio sources with arcsec or subarcsec
spatial resolution (e.g., T.Venturi, this proceeding).
The radio studies suggested the basic modelling of radio jets and
hot spots.
They provided evidence for relativistic motions of 
the radio jets from pc to kpc distances from the nucleus 
(e.g., Garrington et al. 1988; Bridle et al. 1994).
The study of the polarization from jets and hot spots have suggested
the presence of shocks and/or strong interactions with the surrounding
IGM/ICM in which magnetic field amplification and
particle reacceleration can take place.
Finally, the study of the spectral synchrotron ages, combined with 
the direct (or statistical) measurement of the advancing motion of
the radio lobes/hot spots have allowed a first order estimate of the 
dynamical age of extragalactic radio sources.
This in turn allowed the measurement of the
jet kinetic power under the assumption of minimum energy conditions
(Rawlings \& Saunders 1991).
 
Although the improvement of the radio telescopes and interferometers,
and the advent of the future radio instruments 
(e.g., SKA) will allow to address 
a number of additional/substantial improvements in our
understanding of the physics of radio sources, 
a multiwavelength approach is by far
the most efficient tool to provide the next step in this topic.
This is due to the {\it synchrotron degeneracy} (Sect.3.5.1).

{\it OPTICAL OBSERVATIONS}:
although the search for optical emission from radio
jets and hot spots has a long history (e.g., Saslaw et al. 1978;
Simkin 1978; Crane et al. 1983), 
relatively few jets and hot spots have been detected 
as sources of optical emission so far (Tab.2, Tab.3,
Meisenheimer et al. 1997).
This is not only due to the power law decay with frequency
of the synchrotron spectrum emitted from these regions, but also
due to the presence of breaks and/or exponential cut-offs in the
synchrotron spectrum below the optical band (Sect.3.3).
The advent of the {\it Hubble Space Telescope} (HST) and more recently,
of the 10 mt. generation of ground based telescopes 
(e.g., VLT, {\it Gemini}),
has considerably improved the possibility to detect and study the
optical emission from these regions.

{\it X--RAY OBSERVATIONS}:
the study of X--ray emission from jets and hot spots has been
considerably improved by the recent 
advent of the {\it Chandra} observatory.
Before {\it Chandra} only a few cases of X--ray counterparts of
radio jets and hot spots was discovered (see Tab. 2 and 3).
The
most spectacular result being the famous ROSAT HRI detection of both 
hot spots of the powerful radio galaxy Cygnus A (Harris et al. 1994).
It was immediately clear from these past observations 
that the detected emission was of non--thermal
origin with the best interpretation provided by
synchrotron and SSC mechanisms
under approximate minimum energy conditions.
Likewise {\it Chandra} is really providing a
significant progress on the study of the X--ray emission
from jets and hot spots (see Tab. 2 and 3).
Although the analysis of the
increasing number of successful detections of
X--ray emission from compact hot spots and
jets has unambiguously
confirmed the non--thermal nature of the X--rays
from these sources, it is not clear whether the SSC and
synchrotron model
can provide or not a general interpretation of
the data (e.g. Harris 2001). In addition the possibility to derive
spectral analysis of the X--ray counterparts allows, for the first time,
to constrain the spectrum of the emitting electrons in these regions.

In this Section we especially focus on the information on the low
energy end and high energy end of the electron spectrum 
which are becoming available with the most
recent multiwavelength studies.

\subsubsection{Constraining the LOW energy end of 
the electron spectrum}

As discussed in Sect.3.4, information on the low energy
end of the electron spectrum can be 
provided by the detection of optical SSC fluxes from
the hot spots and by the detection of X--ray emission via 
IC/CMB from the jets.

{\em a) Optical SSC emission from hot spots -- 3C 295--N 
and 196--N -- }:
The northern hot spot of 3C 295 has been recently
detected in the B--band with the HST telescope
(Harris et al. 2000).
Taking into account the radio, optical and X--ray data, 
Brunetti (2000) has shown that the radio to optical emission
is not easily accounted for by 
a simple synchrotron model, whereas
a synchrotron plus SSC model can account very well for the
broad band spectrum with the radio matched by the
synchrotron radiation, 
and the optical and X--rays matched by
the SSC (Fig.15).
\begin{figure}
\resizebox{\hsize}{!}{\includegraphics{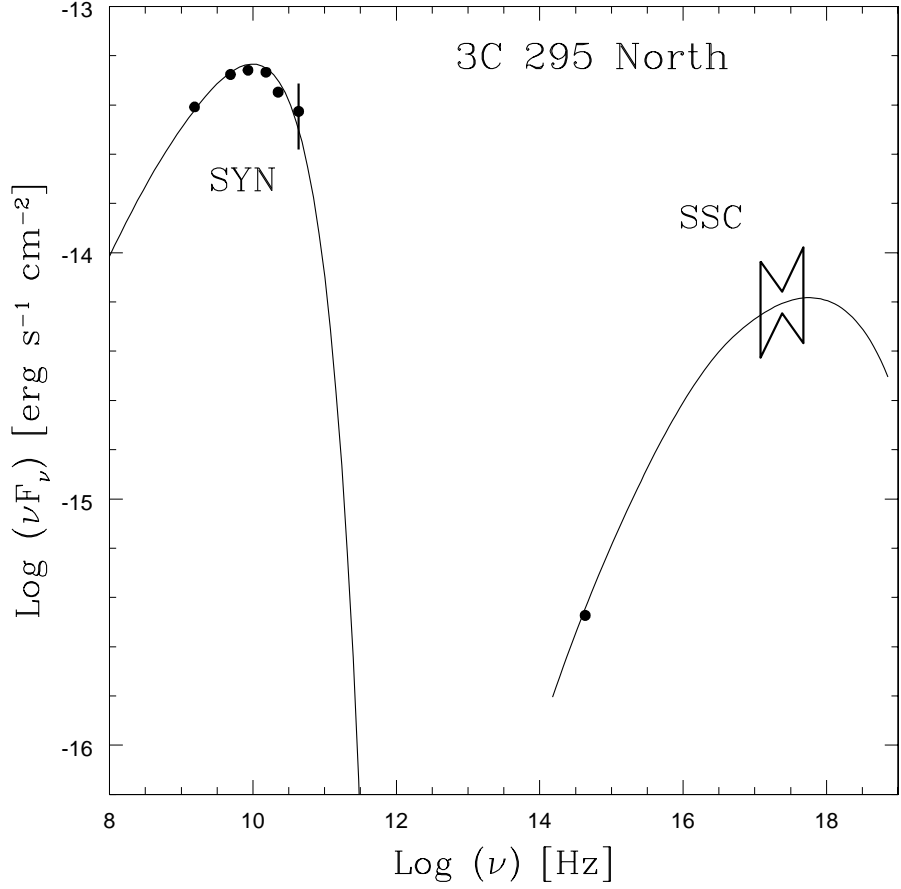}}
\caption{The radio to X--ray data points of
the northern hot spots of 3C 295 are fitted with
a synchrotron and SSC model: 
the SSC accounts for both optical and X--ray data.}
\end{figure}
The synchrotron radiation at $\geq$ GHz frequency is emitted
by $\gamma > 1000$ electrons (Sect.3.4), whereas the SSC optical
radiation is emitted by $\gamma \sim 500$ electrons.
In particular, the model in Fig.15 is calculated assuming that the
spectrum of the synchrotron emitting electrons can be
extrapolated at lower energies ($\gamma_{\rm low} < 500$) and the 
data constrained the low energy break 
in the electron spectrum (if any)
at $\gamma_{*} < 800$ (Sect. 3.3).

The optical counterpart of the northern hot spot
of 3C 196 
has been recently discovered by Hardcastle (2001)
with the HST telescope.
The high frequency radio fluxes show a prominent steepening 
so that a synchrotron model accounting for the radio spectrum
is too steep and falls well below the optical flux.
As in the case of 3C 295-N, a viable explanation for the optical
emission is provided by the SSC mechanism under the assumption 
of approximate equipartition conditions in the hot spots
(Hardcastle 2001).
As in the case of 3C 295, the detection of optical SSC emission 
from the hot spot of 3C 196 points
to the presence of $\gamma \sim 500$ electrons in the hot spot
volume without a significant flattening of the electron
spectrum at these energies.

\begin{table}[htb]
\begin{center}
\caption{High power objects}
\begin{tabular}{lllllll}
\hline
\hline
Name & z & Type &Assoc. & Assoc. & Chandra & Reference\\
     &   &      & Radio  & Optical&         &          \\
\hline
3C 123  & 0.2177  & RG  & HS    & N & Y &  Ha01    \\
Pictor A& 0.0350  & RG  & HS    & Y & N &  W01    \\
        &         &     & Knots & N & Y &      \\
PKS 0637-752& 0.653& FSQ& Knots & Y & Y &  C00,T00,Sc00,Ce01 \\	
3C 179  & 0.846   & FSQ & Knot  & N & Y &  S02    \\
        &         &     & HS    & N & Y &      \\	
3C 207  & 0.684   & FSQ & Knot  & N & Y &  B02b    \\
        &         &     & HS    & N & Y &      \\
Q 0957+561& 1.41  & FSQ & Knots & N & Y &  C02    \\
PKS 1127-145&1.187& FSQ & Knots & N & Y &  Si02    \\
PKS 1136-135&0.554& FSQ & Knots & Y & Y &  S02    \\
4C 49.22& 0.334   & FSQ & Knots & Y & Y &  S02    \\
3C 273  & 0.1583  & FSQ & Knots & Y & N &  M01,S01    \\
4C 19.44& 0.720   & FSQ & Knots & Y & Y &  S02    \\
3C 295  & 0.45    & RG  & HS    & Y & Y &  H00,B01b    \\
3C 351  & 0.3721  & SSQ & HS    & Y & Y &  B01c    \\
3C 390.3& 0.0561  & RG  & HS    & Y & N &  P97,H98    \\
Cyg A   & 0.0560  & RG  & HS    & N & N &  H94,W00    \\
\hline
\hline
\end{tabular}
{\bf References}: Ha01=Hardcastle et al. 2001a, W01=Wilson et al. 2001,
C00=Chartas et al. 2000, T00=Tavecchio et al. 2000,
Sc00=Schwartz et al. 2000, Ce01=Celotti et al. 2001,
S01=Sambruna et al. 2002, B02b=Brunetti et al. 2002b,
C02=Chartas et al. 2002, Si02=Siemiginowska et al. 2002,
M01=Marshall et al. 2001, S01=Sambruna et al. 2001,
H00=Harris et al. 2000, B01b=Brunetti et al. 2001b,
B01c=Brunetti et al. 2001c, P97=Prieto 1997,
H98=Harris et al. 1998, H94=Harris et al. 1994, W00=Wilson et al. 2000.
See http://hea-www.harvard.edu/XJET/index.html for an updated
list.
\end{center}
\end{table}

{\em b) X--ray emission from radio jets -- external IC scattering --}:
one of the most impressive results from {\it Chandra} 
is the unexpected high detection rate of the radio jets in the powerful
radio sources (Tab.2).
The first object with a prominent X--ray jet discovered by {\it Chandra}
was the flat spectrum quasar PKS 0637-752 at a redshift of z=0.653
(Chartas et al. 2000; Schwartz et al. 2000). 
The combined radio, optical and X--ray data exclude the possibility of
synchrotron X--ray emission.
In addition 
Tavecchio et al.(2000) and Celotti et al.(2001) 
have shown that an SSC origin of the X--rays 
should require huge departures from equipartition,  
and an extremely high kinetic power of the jet.
These authors were the first to successfully 
make use of the IC scattering of CMB photons
by the relativistic electrons of the jet
(a mechanism previously considered only for the
jets on pc scales, e.g., Schlickeiser 1996) 
to reproduce the X--ray
data of this object.
In order to successfully fit the X--ray spectrum with this model
they derive a
high relativistic velocity ($\Gamma_{\rm bulk} \sim 10$)
of the jet up to hundreds of kpc distance from the nucleus.
In order to maintain such velocities,  
small radiative efficiencies in the jets are required with most  
of the energy extracted from the central black hole stored
in the bulk motion of the plasma.
In addition, as pointed out by Ghisellini \& Celotti (2001), 
it can be considered that the radiative efficiency
of the jet decreases with increasing 
the jet luminosity.

Additional evidences in favour of the IC/CMB 
and thus that high relativistic velocities are maintained by
the jets up to tens or hundreds of kpc from the nucleus, 
is coming from other objects (e.g., Sambruna et al. 2002;
Tab.2).
In Fig.16a-c we report a compilation of radio to X--ray
spectral energy distributions of some of these objects.

\begin{figure}
\resizebox{\hsize}{!}{\includegraphics{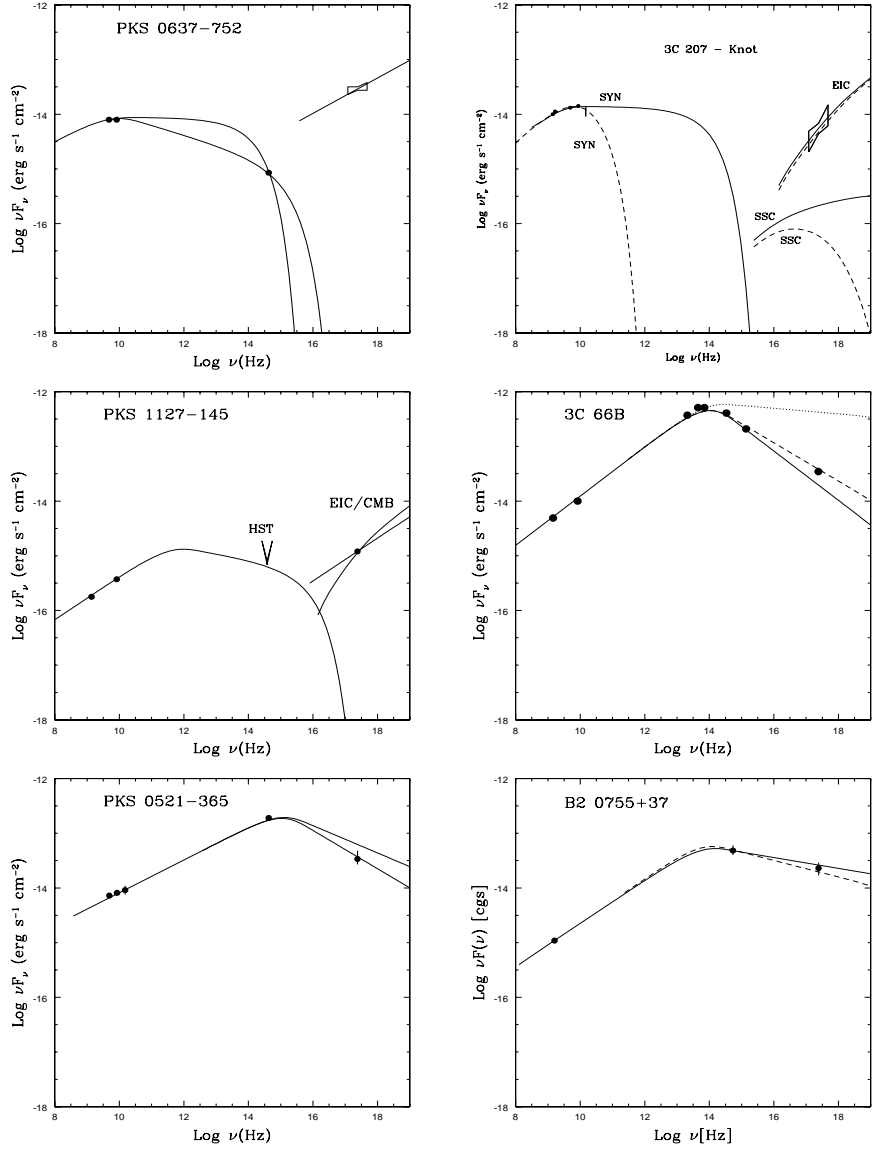}}
\caption{
A compilation of SEDs of X--ray jets
fitted with synchrotron and IC models.
{\bf Panel a)--c)}: SEDs in the case of jets of 
high power radio objects.
Radio and optical data are fitted with synchrotron models,
while X--ray data are accounted for via IC/CMB models.
The SSC predictions are also reported in the case
of 3C 207 (Panel b).
{\bf Panel d)--f)}: SEDs in the case of jets of
low power radio objects.
Radio, optical and X--ray data can be fitted with 
a synchrotron model which includes a non standard 
electron transport or adiabatic losses
in the post--shock region.
Standard synchrotron models (e.g., dotted line in 3C 66B,
upper curves in PKS 0521-365 and B2 0755+37) 
cannot well account for the data. 
}
\end{figure}

One of the most striking cases is the X--ray knot of the
quasar 3C 207 at a redshift z=0.684 (Brunetti et al. 2002b).
The resulting X--ray spectrum is considerably harder than the
radio spectrum ($\Delta \alpha \sim 0.6$) so that it cannot be 
reproduced by the SSC spectrum even releasing the assumption
of minimum energy conditions in the jet.
On the other hand, as discussed in Sect.3.3,
the electron spectrum of the low energy electrons emitting
X--rays via IC/CMB ($\gamma \sim 100$) might
be harder than that of the higher energy radio synchrotron 
electrons ($\gamma > 1000$), thus providing a natural explanation 
for the difference between X--ray and radio spectrum of this knot.
It should be noted that, despite the poor statistics,
a similar difference between radio and X--ray spectrum is also 
found in the jet of PKS 1127-145, which is the most luminous
IC/CMB jet discovered so far (Siemiginowska et al. 2002).

If the IC/CMB interpretation is correct for these X--ray jets,
then, for the first time,  
the modelling of the radio to X--ray spectrum allows 
the low energy end of the electron spectrum
to be constrained in the regions where these electrons 
are (re)accelerated.
These studies, however, are relatively complicated as the energy
of the electrons giving the observed X--rays (Sect.3.4)
depends on both the Lorentz factor of the
bulk motion, 
and on the angle between the jet velocity and the line 
of sight.
In the case of 3C 207 it can be shown that 
for substantial boosting (i.e., $\Gamma_{\rm bulk}>4$ and 
$\theta < 10^o$) $\gamma_{\rm low}$\ltsim 50$ \leq \gamma_*$,
whereas in the case of PKS 0637-752 $\gamma_{\rm low}$\ltsim 30.

\subsubsection{Constraining the HIGH energy end of the electron
spectrum}

{\em a) synchrotron optical emission from radio hot spots}:
As already discussed in Sect.3.4, synchrotron optical
emission from radio hot spots is mainly due to 
$\gamma \geq 10^5$ electrons, which are probably
close to the high energy end of the spectrum of the electrons 
accelerated in these regions.
These electrons have a radiative life time about 300 times
shorther than that of the electrons emitting the synchrotron
radio spectrum of the same hot spots.
Hence the optical
detection of hot spots generally implies 
the in situ production of such energetic electrons 
(e.g., Meisenheimer et al., 1989). 
An important confirmation that the optical emission
from the hot spots is of synchrotron nature
is provided by the detection of optical linear 
polarization in a number of cases 
(3C 33: Meisenheimer \& R\"oser, 1986;
3C 111, 303, 351, 390.3: Lahteenmaki \&
Valtaoja 1999; Pictor A West: Thomson et al. 1995).
So far there are only about 15 hot spots detected in the
optical band (see Gopal-Krishna et al. 2001 and ref. therein)
and the radio to optical spectrum of these hot spots is
well fitted by synchrotron radiation emitted by electrons
accelerated in a shock region (e.g., Meisenheimer et al. 1997).
In Fig.17 we report the radio to optical data of a few 
representative cases fitted with synchrotron models.

In principle,
if this scenario is correct, 
the theory of shock acceleration allows us 
to get an independent estimate of the field strength at the hot spot
by measuring the break and cut--off frequencies of the synchrotron
spectrum and the hot spot
length (e.g., Meisenheimer et al., 1989).
This allows the 
estimate of
both the maximum energy of the emitting electrons, 
and of the acceleration efficiency of the shock.
In general, the estimated magnetic field strength is 
consistent with that estimated under minimum energy
conditions within a factor of 2--3
(e.g., Meisenheimer et al., 1997).
With these values, we have that the Lorentz factors of
the electrons at the cut--off is in the range 
$\gamma_{\rm c}=10^5-10^6$, 
and that 
the acceleration time in the shock region is
in the range $\sim 10^2-10^3$yrs. 

\begin{figure}
\resizebox{\hsize}{!}{\includegraphics{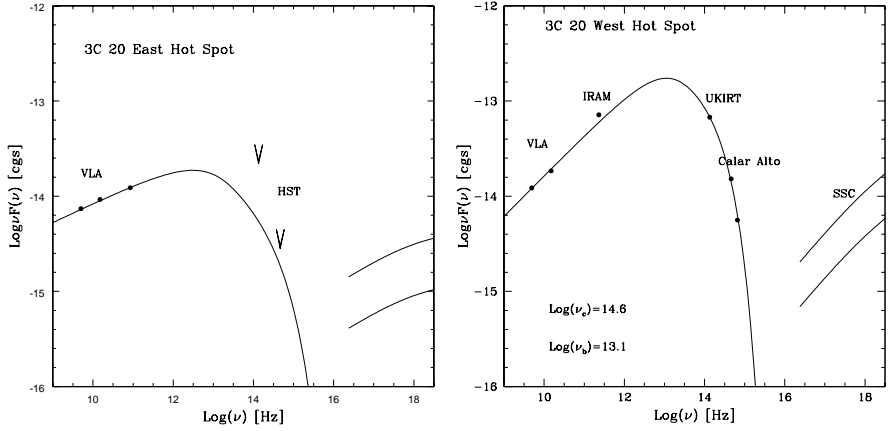}}
\caption{
Synchrotron fits to the 
radio -- optical SEDs of the radio hot spots 
3C 20 East and West are reported.
The spectrum of the emitting electrons 
is calculated under shock acceleration assumptions
(Sect.3.3, Figs.3-4).
In the case of 3C 20 West the values of the break and
cut--off frequency are also given.
The expected X--ray emission from both the hot spots
due to SSC process are reported
under equipartition conditions (lower curves) and assuming
a magnetic field 2 times smaller than 
the equipartition (higher curves).
}
\end{figure}

An alternative scenario to that of the shock
acceleration might be an extremely efficient
transport -- minimum energy losses -- of the ultra relativistic
electrons all the way from the core to the hot spots
(e.g., Kundt \& Gopal-Krishna 1980). 
Based on the evidences for relativistic jet bulk motion out
to 100--kpc scales, Gopal-Krishna et al. (2001) 
have recently
reconsidered a minimum loss scenario in which the
relativistic electrons, accelerated in the central active nucleus,
flow along the jets losing energy only due to the inescapable
IC scattering of CMB photons.
Under these assumptions, comparing the electron radiative 
life time with the travel time to the hot spots, these authors 
find that in situ electron re--acceleration is in general not
absolutely necessary to explain the optical synchrotron
radiation from the hot spots.
In the framework
of this minimum loss scenario, in Fig.18
we report the maximum
distance that synchrotron optical electrons 
can cover as a function of the velocity of the jet flow, and
for two 
different magnetic field strengths in the hot spot region.
\begin{figure}
\resizebox{\hsize}{!}{\includegraphics{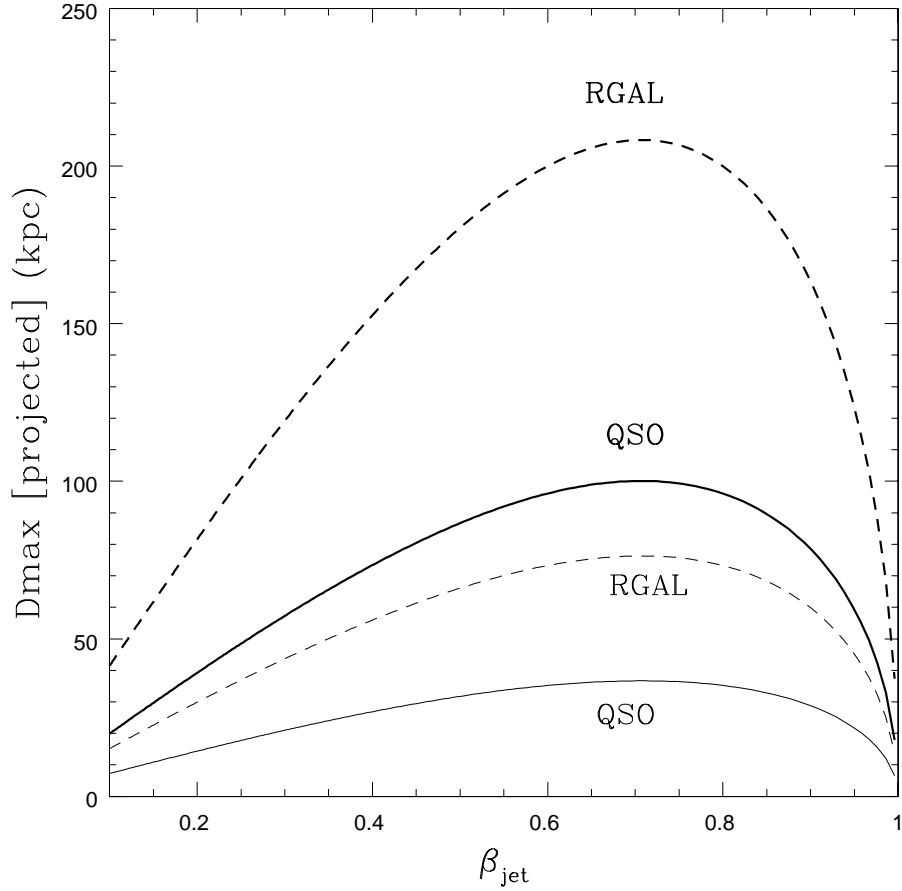}}
\caption{
The maximum distance (projected on the plane of the
sky) that electrons emitting synchrotron 
radiation at $\nu = 5 \cdot 10^{14}$Hz 
can cover is reported as a 
function of the velocity (v/c) of the jet.
The calculation are performed for different inclination angles 
of the jet: quasar--like (solid lines) and radio galaxy--like 
(dashed lines).
The results are shown in the case of z=0.2 (thin lines) and
z=0.5 (thick lines).
}
\end{figure}
Such distance is in general less
than 100 kpc except in the case in which the magnetic field
strength in the hot spot is very large.
On the other hand, a number of optically detected radio
hot spots are found at larger distances from the nucleus.
In addition, the double hot spot in 3C 351 represents a clear 
counter example to the minimum energy loss scenario.
Indeed, in this case the hot spot magnetic field ($B \leq 100 \mu$G)
is constrained matching the X--ray flux by the SSC process, 
and the distance of the hot spots from the nucleus 
is $> 180$ kpc.
As both the hot spots emit synchrotron radiation at optical
wavelengths, in situ reacceleration appears to be 
inescapable (Brunetti et al. 2001c).

{\em b) synchrotron X--ray emission from jets}:
One of the most interesting findings of {\it Chandra} is
that X--ray jets are relatively common also in the
case of low power radio sources (Worral et al., 2001).
These X--ray jets are usually interpreted as synchrotron
emission from the very high energy end ($\gamma \sim 10^7$) 
of the electron population.
Such interpretation is mainly supported by the observed radio
to X--ray spectral distributions.
In addition, it is 
supported by the fact that, contrary to the case of high
power radio sources, the jets of low power objects are 
believed to move at sub/trans--relativistic speeds
at kpc distances from the nucleus (e.g., T.Venturi, this
proceedings and ref. therein) and thus the X--ray emission
from IC/CMB is expected to be negligible.
If so, the jets of low power radio sources can be considered
laboratories to study the electron acceleration.
Indeed, combining radio, optical and X--ray data it is  
possible to study the synchrotron spectrum from radio to 
X--ray frequencies and thus to sample the spectrum of the
emitting electrons over more then 4 decades in energy.
In particular, relatively deep multiwavelength observations
with adequate frequency coverage 
of a few objects (M 87: Boehringer et al. 2001; 
3C 66B: Hardcastle et al. 2001;
PKS 0521-365: Birkinshaw et al. 2002) 
have shown that the radio to X--ray
spectrum can be well fitted by 
a double power law model of slope 
$\alpha$ and $\alpha + (0.7-0.9)$ in 
the radio and in the optical to X--ray band, respectively.
If further confirmed, this point is crucial as it generates 
problems in the interpretation of the data with 
acceleration models including standard electron diffusion
in the post shock region (e.g. Bell, 1978a,b; 
Heavens \& Meisenheimer, 1987) which, indeed, 
would predict a steepening of the synchrotron spectrum of only 
0.5 in the optical to X--ray band.

More recently Dermer \& Atoyan (2002) have proposed that 
synchrotron emission can successfully fits the X--ray 
data also in the case of some of the detected X--ray jets of
high power radio sources usually interpreted
via IC/CMB scattering.
These authors have investigated 
the evolution of the spectrum of electrons accelerated
up to very high energies ($\gamma > 10^8$) under the 
hypothesis that the radiative losses of the electrons
are largely dominated by IC scattering rather than 
synchrotron. 
Under these conditions, if the photon energy density 
in the jet frame is dominated
by boosted CMB photons (as in the case of the IC/CMB
process), the energy dependence of the 
IC losses of the electrons with $\gamma > 10^7$ 
changes due to the effect of the Klein--Nishina cross section
and the radiative losses for these electrons result alleviated
(Fig.1). 
It can be shown that, under these conditions, 
the spectrum of the electrons may become harder
for $\gamma > 10^7$ 
and that the resulting synchrotron 
emission may present a bump in the X--ray band 
similarly to that observed by {\it Chandra}.
Assuming a transverse velocity structure 
in the jets, with a fast central spine surrounded with a 
boundary layer with a velocity shear (Sect. 4.3.5), 
it has been proposed that 
turbulence may also accelerate high energy 
electrons at such boundary layers 
(Owstroski, 2000; Stawarz \& Ostrowski, 2002a).
If this happens, X--ray synchrotron radiation from large 
scale jets is expected and it may account for some of
the observed {\it Chandra} jets 
(Stawarz \& Ostrowski, 2002b).

{\em c) synchrotron X--ray emission from hot spots ?}:

The effect of electron radiative cooling and the
presence of a high energy cut--off in the electron spectrum
produce an abrupt steepening of the
spectrum of the hot spots below that extrapolated from the 
lower frequency power--low.
This makes X--ray
detection of synchrotron emission very difficult.
In addition the photons emitted by
competing processes particularly efficient
at high frequencies (e.g., SSC) might completely
hide those contributed by
the synchrotron emission.
So far, there are only two relatively secure
cases of hot spots in which the synchrotron spectrum 
is given by a power law from the radio to the 
UV or even X--rays (3C 303 : Keel 1988, Meisenheimer et al. 1997
; 3C 390.3 : Prieto 1997, Harris et al. 1998),
indicating the continuation of the synchrotron
spectrum at higher frequencies.
An immediate implication is the presence of relativistic 
electrons with $\gamma \geq 10^7$ which, due to
their short life time (considering typical hot spots' 
magnetic field strength $\geq 100 \mu$G), require very efficient 
acceleration processes (acceleration time $\leq 100$ yrs) and/or
magnetic field strengths in the acceleration regions 
well below that calculated under equipartition conditions.

\subsubsection{Energetics: 
X--ray SSC emission from radio hot spots in FR II}

Until the advent of {\it Chandra} clear evidence for
SSC emission had only been detected in the case of the 
hot spots of Cygnus A (Harris et al., 1994) in which case
the magnetic field results close to the
equipartition value.

{\it Chandra} has enabled significant progress in this field,
with a number of successful detections in the first
three years of observations (3C 295: Harris et al. 2000;
Cyg A: Wilson et al. 2000; Pictor A: Wilson et al. 2001;
3C 123: Hardcastle et al. 2001; 3C 351: Brunetti et al. 2001c).
 
In the majority of the detected hot spots
(Cygnus A--W and E, 3C 295--N, 3C 123) the magnetic fields derived 
comparing the radio and X--ray fluxes (Sect.3.5.2) 
result within a factor of 2 from the equipartition
value. This has further motivated 
the usually adopted assumption
of approximate 
equipartition between magnetic field and electron
energy densities. 

On the other hand, in the case of the double northern
hot spots of 3C 351 (J and L components),  
if the SSC interpretation is correct, the magnetic field
would result in both cases from a factor 3.5 to 5 smaller than the
equipartition value
(in case of ordered or isotropic field configuration, respectively). 
Here we stress that this departure from
equipartition implies an energy density of the electrons in the
hot spots a factor $> 60$ larger than that of the magnetic field.
Such a relatively strong departure from equipartition in these
hot spots is further suggested 
by the modelling of the broad 
band synchrotron spectrum (Fig.19).
Indeed, the magnetic field intensity,  
derived combining the optical 
synchrotron cut--off and break frequencies with the hot spots' lengths 
along the jet direction, lead to an independent estimate of the 
magnetic field strengths which is 
in good agreement with those 
obtained with the SSC argument (Brunetti et al. 2001c).
\begin{figure}
\resizebox{\hsize}{!}{\includegraphics{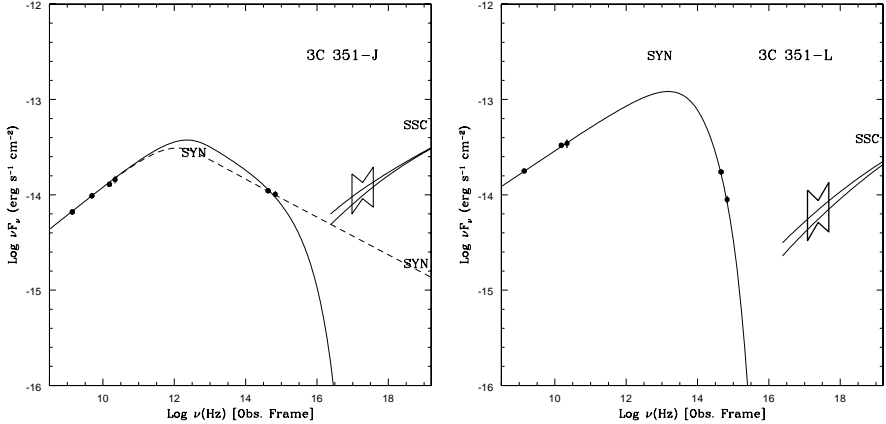}}
\caption{
The radio, optical and X--ray data of the
hot spots 3C 351--J (left panel) and 3C 351--L (right panel)
are reported together with the synchrotron and SSC spectra.
The spectrum of the emitting electrons is calculated assuming
shock acceleration (Sect.3.3, Figs.3-4).
Radio and optical fluxes are accounted for by synchrotron
emission, whereas the X--rays are matched via SSC.
In both cases, the reported SSC models are calculated assuming 
a substantial departure from equipartition conditions (see text
for details). 
}
\end{figure}

A particular intriguing case is the west hot spot of Pictor A
(Wilson et al. 2001).
The synchrotron spectrum shows an abrupt cut--off clearly
indicated by a large number of optical data points 
and it falls orders of magnitude below the X--ray flux.
On the other hand, the SSC interpretation would require 
a magnetic field strength about a factor 14
below the equipartition value to match the observed
X--ray flux. In addition, the {\it Chandra} X--ray spectrum
of the hot spot is relatively steep ($\alpha \simeq 1.1$) and it
is poorly fitted by the spectrum expected in case
of SSC emission ($\alpha \simeq 0.74$).

\begin{table}[htb]
\begin{center}
\caption{Low power objects}
\begin{tabular}{lllllll}
\hline
\hline
Name & z & Type &Assoc. & Assoc. & Chandra & Reference\\
     &   &      & Radio  & Optical&         &          \\
     \hline
     3C 31   & 0.0167  & RG  & Knots & N & Y &  Ha02    \\
     B2 0206+35&0.0368 & RG  & Knots & N & Y &  Wo01    \\
     3C 66B  & 0.0215  & RG  & Knots & Y & Y &  Ha01    \\
     3C 120  & 0.0330  & RG  & Knot  & N & N &  H99    \\
     B2 0755+37& 0.0428& RG  & Knots & Y & N &  Wo01    \\
     3C 270  & 0.00737 & RG  & Knots & Y? & Y & Ch02    \\
     M 87    & 0.00427 & RG  & Knots & Y & N &  Bi91,M02,W02    \\
     Cen A   & 0.001825& RG  & Knots & ? & Y &  F91,K02    \\
     3C 371  & 0.051   & BL  & Knots & Y & Y &  P01    \\
     \hline
     \hline
     \end{tabular}
     {\bf References}: Ha02=Hardcastle et al. 2002,
     Wo01=Worrall et al. 2001, Ha01=Hardcastle et al.
     2001b, H99=Harris et al. 1999, Ch02=Chiaberge et al. 2002,
     Bi91=Biretta et al. 1991, M02=Marshall et al. 2002,
     W02=Wilson \& Yang 2002, F91=Feigelson et al. 1991,
     K02=Kraft et al. 2002, P01=Pesce et al. 2001.
     See also http://hea-www.harvard.edu/XJET/index.html for an
     updated list.
     \end{center}
     \end{table}

\subsubsection{Multiple electron populations in jets and hot spots ? :
Pic A west and 3C 273 jet}

One possibility to match the radio, optical and X--ray data of
Pictor A west is to have two electron populations (Wilson et al.
2001).
The first population of electrons (with injection
spectral index $\delta \simeq 2.5$) 
is assumed to be in a region of relatively strong magnetic field
(e.g. of the order of the equipartition field) 
and would be the responsible
for the observed radio to optical spectrum via synchrotron emission.
As already noticed, such a population would produce a SSC emission
about two orders of magnitude below the observed X--ray flux.
Thus it might be assumed a {\it second population} 
of relativistic 
electrons in the hot spot 
(with injection spectral index $\delta \simeq 3.3$), spatially separated 
from the first population and in low field regions.
This {\it second population} 
would emit negligible synchrotron radiation
and it may produce efficient IC scattering of the synchrotron 
radio--optical photons (from the first population) 
matching the observed X--ray spectrum.
Alternatively, the electron spectrum of the
second population might extend up to very high
energies (with $\nu_{\rm c}> 10^{17}$Hz)
matching the X--rays via synchrotron radiation.
In this last
case a value of the injection spectral index $\delta \simeq 2.15$
and a break frequency in between $10^{11}$ and $10^{17}$Hz is required
in order to not overproduce the synchrotron spectrum
emitted by the first population (Wilson et al. 2001).
Both these possibilities are {\it ad hoc} and, so far,
they are not well
physically motivated so that the case of Pictor A west remains 
poorly understood and future radio and {\it Chandra} observations 
are still required.

A second interesting case of possible multiple electron
populations is the well studied jet of 3C 273.
Recently, Marshall et al.(2001) and Sambruna et al.(2001) 
performed a detailed radio, optical (HST) and 
X--ray ({\it Chandra}) study of this jet.
These authors provide different interpretations of the stronger 
knots in this jet.
In particular, Sambruna et al.(2001) have suggested an IC/CMB
interpretation for the X--ray spectrum of the knots 
since a synchrotron model from a single electron population
cannot fit the radio to X--ray spectra.
On the other hand, making use of a different data set, 
Marshall et al.(2001) were able to fit the
observed spectra with a single synchrotron model.
The most important difference in the two data sets is 
given by the
different slopes of the optical spectra; 
additional observations and more detailed data analysis
are probably requested to understand this discrepancy.
A detailed optical--UV study of the jet of 3C 273 has been
performed by Jester et al.(2001). These authors have measured the
optical--UV spectral index along the jet discovering that it is
only slowly changing without showing a clear trend along the jet.
This has been interpreted as the signature of 
continuous reacceleration processes (Fermi I and II -- like)
of the relativistic electrons active along the jet 
which, in principle, 
might yield multiple electron populations or
a single electron population with a complex spectrum.
It should be noticed that, indeed, Sambruna et al.(2001) 
do not exclude the possibility to fit the radio to X--ray spectrum 
of the knots with synchrotron emission by {\it ad hoc} multiple 
(or complex) electron populations.
A further step forward in the study of the jet of 3C 273 has been 
recently obtained by Jetser et al.(2002).
These authors discovered the presence of an UV excess in the bright
knots with respect to that expected from 
standard synchrotron models.
This strengthens the possibility
of a complex spectrum of the electrons or the presence of coexisting
multiple electron populations.
Jester et al.(2002) have also pointed out that
the UV excess in the knots A and D2+H3
is consistent with the contribution due to the
extrapolation at UV frequencies of the X--ray spectrum
thus suggesting a common origin for 
the UV excess and for the X--rays.
At the light of these results, we might conclude that the 
UV to X--ray spectrum can be produced 
by IC/CMB radiation emitted by 
the electron population giving the 
radio to optical emission via synchrotron process 
(as in Sambruna et al.2001).
On the other hand, a synchrotron origin for the UV to X--ray 
spectrum of the bright knots 
cannot be excluded if a second - high energy -
electron population
is assumed (Jester et al. 2002).
Additional multifrequency observations are required to test the
different scenarios.

\subsubsection{Velocity structure in radio jets}

Komissarov (1990) suggested that the emission minima observed near
the starting point of some FR I jets could result from Doppler dimming 
and that the appearance of the jet might be due to the presence of 
a slow moving boundary {\it layer} which would be 
less dimmed than the faster internal {\it spine}.
Laing (1993) made the connection to polarization structure of FR I
jets proposing a two component model consisting of an 
internal high velocity {\it spine} containing a magnetic field 
which has no longitudinal component but is otherwise random and
a lower velocity external {\it layer} with an entirely longitudinal
field structure.
If FR I jets have field and velocity structure of this type and
if they are launched at relativistic speeds decelerating 
away from the nucleus, the emission of the {\it spine} and {\it layer}
components will suffer different effects due to beaming. 
Consequently, which component would
dominate the emission properties of the 
jets depends on the angle with the line of sight 
and on the distance from the nucleus.
This model provides a natural explanation for the tendency 
of the apparent magnetic field direction to be longitudinal close
to the core and transverse further out.
Detailed applications of models with velocity structure
have been performed in the case of 
individual radio galaxies (e.g., 3C 31: Laing 1986;
3C 296: Hardcastle et al. 1997)
and also for a small sub--sample selected from the 
B2 sample (Laing et al. 1999).
More recently, Chiaberge et al.(2000) have explored 
the viability of the unification
of BL Lacs and FR I radio galaxies by comparing the core emission of
radio galaxies with those of BL Lacs of similar extended radio power
in the radio-optical luminosity plane.
In agreement with the Komissarov \& Laing findings, 
these authors conclude that velocity structures in the jet are 
necessary to reconcile the observations with the unification scheme.

A possibility to study the velocity structure of radio jets
(of both high and low power objects) on kpc scales is to compare 
their radio and X--ray emission properties.
Assuming a transverse velocity structure,
the emission of the jets pointing in the direction of
the observer should be dominated by the contribution of
the fast moving {\it spine}, whereas that from a misaligned jet 
should be dominated by the emission of the slow moving {\it layer} 
which is less dimmed by transverse Doppler boosting.
In the case of high power radio objects, 
this scenario might be easily tested by {\it Chandra}.
Indeed, these jets are highly collimated up
to tens or hundreds kpc distance from the nucleus and
thus indicating that their velocity structure (if any) should 
be preserved on large scales.
As stated in Sect.4.3.1, the X--ray emission from the
jets of core dominated quasars indicate a highly relativistic
motion up to tens of kpc distance from the core.
These velocities might be associated with a fast {\it spine} 
with a low radiative efficiency.
On the other hand, future {\it Chandra} observations of the jets 
of the misaligned parent population (FR II radio galaxies) might
constrain the contribution from any slow {\it layer}.

\section{SUMMARY}

In this contribution we have tried to summarize new insights
on the physics of extragalactic radio sources from
recent studies based on combined observations in different bands.
We concentrated ourselves on the case of the non--thermal emission
produced in kpc--scale regions (radio lobes,
kpc--scale jets, hot spots) and on the spectrum and energetics of
the emitting relativistic electrons.

We have shown that the spectrum of the
relativistic electrons in radio sources can be approximated with
a power law only in a relatively narrow energy range.
In general, the shape of the spectrum depends on the acceleration
mechanisms active in the emitting regions and on the competition
between such mechanisms and both the processes responsible for the
energy losses and the spatial diffusion
of the relativistic electrons.
Measurements of the flux and spectrum produced by
different non--thermal emitting processes in different frequency bands 
can allow to trace the spectrum of the emitting electrons and thus
to constrain the physics of the acceleration in these remote
regions.
Such measurements are possible only now by combining radio, optical and
X--ray observations with arcsec resolution.

Here, a `biased' summary of some of the most promising recent findings:

$\bullet$
{\it Low energy end of the electron spectrum}:
An advance in constraining the energetics associated
to the low energy electrons has coming from 
the new detections of extended X--ray emission from IC scattering 
of nuclear photons in powerful FR II radio galaxies and quasars.
This effect opens a new window on the study of
electrons with $\gamma \sim 100$ in radio lobes
which are invisible in the radio 
band as, in general, they would emit synchrotron radiation in the 
0.01 to 1 Mhz frequency range.
We have stressed that, making use of the classical minimum energy 
formulae, these electrons are not taken into account in the
calculation of the energetics of the radio lobes.
On the other hand, 
the first X--ray detections of IC scattering of nuclear photons 
indicate that the bulk of the energetics of the radio lobes
is contained by these low energy electrons.
Additional evidences that the spectrum of the relativistic
electrons extends down to low energies ($\gamma < 500$)
is provided by the
possible detection of SSC optical emission from radio
hot spots and by the X--ray emission from 
powerful radio jets due to boosted
IC scattering of CMB photons.

$\bullet$
{\it High energy end of the electron spectrum}:
The maximum energy of the electrons in extragalactic radio sources
depends on the balance between acceleration and losses mechanisms.
The discovery of optical synchrotron emission from hot spots
in the last decades has pointed out the presence of efficient
accelerators active in these regions and able to accelerate electrons
up to very high energies ($\gamma \geq 10^5$).
More recently, the discovery of synchrotron X--ray emission from
an increasing number of jets of low power radio galaxies 
suggest the presence of even more energetic electrons 
($\gamma \geq 10^{6-7}$) 
possibly accelerated in low B--field regions. 
These findings and the recent suggestion about 
synchrotron X--ray emission also from jets of powerful sources,
would indicate the presence of extremely efficient accelerators 
in the emitting regions.
The extremely short acceleration time--scales (down to $\sim 10^2$ yrs)
requested by these findings are actually starting to put 
interesting limits on the spatial diffusion coefficient of 
the electrons and thus on the microphysics prevailing
in these regions.
The recent additional evidence for continuous reacceleration of 
relativistic electrons in radio jets might finally suggest the 
presence of energetic
turbulence and thus that part of the kinetic power of
the jets is dissipated into the developing of plasma 
instabilities and in the re--acceleration of relativistic
particles.
This might match with the proposed scenario 
in which a slow {\it layer}, where a fraction of
the kinetic power is dissipated, surrounds a fast moving
{\it spine}.

$\bullet$
{\it Energetics of radio lobes and hot spots}:
A crucial point in the study of the extragalactic 
radio sources is the calculation  
of the energetics associated to these objects.
As we have shown, the advent of {\it Chandra}
makes it possible to extensively apply the IC method and to derive 
the energy density of the relativistic electrons and of the 
magnetic field in the emitting regions.
This has been done in the case of a number of
hot spots and radio lobes : the derived energetics are usually
larger (1 to 30 times) than that calculated with the classical 
minimum energy formulae.
This is a direct consequence of the presence of low energy 
electrons which contain most of the energetics, 
but it is also due to moderate departures from the minimum energy 
conditions found in a number of cases.
Additional statistics is requested to better address this point.
We further 
claim that deep X--ray follow up of the most unambiguous 
detections of diffuse IC emission from radio lobes will probably 
allow to derive the first maps of the magnetic field intensity in 
the extragalactic radio sources providing unvaluable information 
on the prevailing physics.

$\bullet$
{\it Kinematics of the radio jets}:
One of the most interesting findings obtained combining the radio,
optical and {\it Chandra} X--ray data of powerful radio jets is
the recent discovery of highly relativistic speeds of the
jets up to several tens of kpc of distance from the nucleus.
Bulk Lorentz factors 
$\Gamma_{\rm bulk} \sim$ 3--10 up to these distances 
limit the radiative efficiency of
the jets that should be low in order to preserve the kinetic power
of the jet itself.
Again, a scenario with a velocity structured jet with 
a fast {\it spine} surrounded by a slow {\it layer} 
might help to better understand this findings.

\vskip 0.2cm

\noindent
{\it Acknowledgements}
I am grateful to all my collaborators, in particular
to M. Bondi, A. Comastri and G. Setti for help 
and discussions.
I am indebted to F. Mantovani who invited me to
give these lectures and to M. Marcha for a careful 
reading of the manuscript and for useful comments.


\begin{thebibliography}{}  
\bibitem[]{}
Aharonian, F.A., Atoyan, A.M., 1981, {\it Ap\&SS},
{\bf 79}, 321
\bibitem[]{}
Arshakian, T.G., Longair, M.S., 2000, {\it MNRAS}, {\bf 311}, 846
\bibitem[]{}
Bell, A.R., 1978a, {\it MNRAS}, {\bf 182}, 147
\bibitem[]{}
Bell, A. R., 1978b, {\it MNRAS}, {\bf 182}, 443
\bibitem[]{}
Bicknell, G. V., Melrose, D. B., 1982, {\it ApJ}, {\bf 262}, 511
\bibitem[]{}
Birkinshaw, M., Worrall D. M., Hardcastle, M. J., 2002, 
{\it MNRAS}, in press; astro-ph/0204509
\bibitem[]{}
Biretta, J. A., Stern C. P., Harris, D. E., 1991, {\it AJ}, {\bf 101},
1632
\bibitem[]{}
Blandford, R. D., 1986, in {\it Magnetospheric Phenomena in
Astrophysics}, eds. R.I. Epstein \& W.C. Feldman
(New York: AIP), p.1
\bibitem[]{}
Blandford, R. D., Ostriker, J. P., 1978, {\it ApJ}, {\bf 221},
L29
\bibitem[]{}
Blandford, R. D., Eichler, D., 1987, {\it Physics Rep.},
{\bf 154}, 1
\bibitem[]{}
Blasi, P., 2000, {\it ApJ}, {\bf 532}, L9
\bibitem[]{}
Blasi, P., 2001, {\it APh}, {\bf 15}, 275
\bibitem[]{}
Blumenthal, G. R., Gould, R. J., 1970, {\it RvMP}, {\bf 42},
237
\bibitem[]{}
B\"ohringer, H., Belsole, E., Kennea, J., et al., 2001,
{\it A\&A}, {\bf 365}, L181
\bibitem[]{}
Borovsky, J. E., Eilek, J. A., 1986, {\it ApJ}, {\bf 308}, 929
\bibitem[]{}
Bridle, A. H., Perley, R. A., 1984, {\it Ann. Rev. Astr. Ap.},
{\bf 22}, 319
\bibitem[]{}
Bridle, A. H., Hough, D. H., Lonsdale, C. J., Burns, J. O.,
Laing, R. A., 1994, {\it AJ}, {\bf 108}, 766
\bibitem[]{}
Brunetti, G., 2000, {\it Astroparticle Physics}, {\bf 13}, 107
\bibitem[]{}
Brunetti, G., Setti, G., Comastri, A., 1997,
{\it A \& A}, {\bf 325}, 898
\bibitem[]{}
Brunetti, G., Comastri, A., Setti, G., Feretti, L., 1999,
{\it A \& A}, {\bf 342}, 57
\bibitem[]{}
Brunetti, G., Setti, G., Feretti, L., Giovannini, G., 2001a,
{\it MNRAS}, {\bf 320}, 365
\bibitem[]{}
Brunetti, G., Cappi, M., Setti, G., Feretti, L.,
Harris, D. E., 2001b, {\it A \& A}, {\bf 372}, 755
\bibitem[]{}
Brunetti, G., Bondi, M., Comastri, A., et al., 2001c, 
{\it ApJ}, {\bf 561}, L 157
\bibitem[]{}
Brunetti, G., Comastri, A., Dallacasa, D., Bondi, M., Pedani, M,
Setti, G. 2002a, {\it Proc. Symposium New Visions of the X-ray
Universe in the XMM-Newton and Chandra Era'}; astro-ph/0202373
\bibitem[]{}
Brunetti, G., Bondi, M., Comastri, A., Setti, G., 2002b,
{\it A \& A}, {\bf 381}, 795
\bibitem[]{}
Celotti, A., Ghisellini, G., Chiaberge, M., 2001, 
{\it MNRAS}, {\bf 321}, L1
\bibitem[]{}
Chartas, G., Worrall, D. M., Birkinshaw, M., et al., 2000, 
{\it ApJ}, {\bf 542}, 655
\bibitem[]{}
Chartas, G., Gupte V., Garmire G., et al., 2002, {\it Apj},
{\bf 565}, 96
\bibitem[]{}
Chiaberge M., Celotti, A., Capetti, A., Ghisellini, G., 2000, 
{\it A\&A}, {\bf 358}, 104
\bibitem[]{}
Chiaberge M., Gilli, R., Macchetto, F. D., Sparks, W. B., Capetti, A.,
2002, {\it ApJ}, submitted; astro-ph/0205156 
\bibitem[]{}
Clark, D. H., Caswell, J. L., 1976, {\it MNRAS}, {\bf 174}, 267
\bibitem[]{}
Cox, A. C., 1999, {\it Allen's Astrophysical Quantities}
(Springer-Verlag)
\bibitem[]{}
Crane, P., Tyson, J. A., Saslaw, W. C., 1983, {\it ApJ}, 
{\bf 265}, 681
\bibitem[]{}
Daly, R. A., 1992, {\it ApJ}, {\bf 386}, L9
\bibitem[]{}
Dermer, C. D., 1995, {\it ApJ}, {\bf 446}, L63
\bibitem[]{}
Dermer, C. D., Atoyan, A. M., 2002,  {\it ApJ}, {\bf
568}, L81
\bibitem[]{} 
Dickel, J.R., 1983, in {\it Supernova remnants and their X-ray 
emission, IAUS}, {\bf 101}, 213
\bibitem[]{}
Eilek, J. A., Hughes, P. A., 1991, in {\it
Beams and Jets in Astrophysics}, eds. P.A.Hughes (Cambridge
astrophysics series),
p.428
\bibitem[]{}
Eilek, J. A., 1996, in {\it Extragalactic radio sources, IAUS},
{\bf 175}, 483
\bibitem[]{}
Eilek, J. A., Arendt, P. N., 1996, {\it ApJ},
{\bf 457}, 150
\bibitem[]{}
Fabian A. C., Crawford, C. S., Ettori, S., Sanders, J. S., 
2001, {\it MNRAS}, {\bf 322}, L11
\bibitem[]{}
Feigelson, E. D., Schreier E. J., Delvaille J. P.,
et al., 1981, {\it ApJ}, {\bf 251}, 31
\bibitem[]{}
Feigelson, E. D., Laurent-Muehleisen, S. A.,
Kollgaard, R. I., Fomalont, E. B., 1995,
{\it ApJ}, {\bf 449}, L149
\bibitem[]{}
Felten, J. E., Morrison, P., 1966, {\it ApJ}, {\bf 146}, 686
\bibitem[]{}
Feretti, L., Giovannini G., 1996, in {\it
Extragalactic radio sources, IAUS}, {\bf 175}, 333
\bibitem[]{}
Ferrari, A., Trussoni, E., Zaninetti, L., 1979,
{\it A\&A}, {\bf 79}, 190
\bibitem[]{}
Garrington, S.T., Leahy, J. P., Conway, R. G.,
Laing, R. A., 1988, {\it Nature}, {\bf 331}, 147
\bibitem[]{}
Ghisellini, G., Celotti, A., 2001, 
{\it MNRAS}, {\bf 327}, 739
\bibitem[]{}
Ginzburg, V. L., 1969,
{\it Elementary Processes for Cosmic Ray Astrophysics}
(Gordon \& Breach Science Publishers: New York)
\bibitem[]{}
Gitti, M., Brunetti, G., Setti, G., 2001,
{\it A\&A}, {\bf 386}, 456
\bibitem[]{}
Gopal-Krishna, Subramanian, P., Wiita, P. J., Becker, P. A.,
2001, {\it A\&A}, {\bf 377}, 827
\bibitem[]{}
Gould, R. J., 1979, {\it A\&A}, {\bf 76}, 306
\bibitem[]{}
Hamilton, R. J., Petrosian, V., 1992, {\it ApJ}, {\bf 398}, 350
\bibitem[]{}
Hardcastle, M. J., 2001, 
{\it A \& A}, {\bf 373}, 881
\bibitem[]{}
Hardcastle, M. J., Alexander, P., Pooley, G. G.,
Riley, J. M., 1997, {\it MNRAS}, {\bf 288}, L1
\bibitem[]{}
Hardcastle, M. J., Worrall, D. M., 2000,
{\it MNRAS}, {\bf 319}, 562
\bibitem[]{}
Hardcastle, M. J., Birkinshaw, M., Worrall, D. M., 2001a,
{\it MNRAS}, {\bf 323}, L 17
\bibitem[]{}
Hardcastle, M. J., Birkinshaw, M., Worrall, D. M., 2001b,
{\it MNRAS}, {\bf 326}, 1499
\bibitem[]{}
Hardcastle, M. J., Worrall, D. M., Birkinshaw, M., Laing R. A., 
Bridle A. H., 2002, {\it MNRAS}, in press; astro-ph/0203374
\bibitem[]{} Harris, D. E., 2001, in ``Particles and Fields in Radio
Galaxies'',
R. A. Laing and K. M. Blundell (Eds.), ASP Conf. Series, in press;
astro--ph/0012374
\bibitem[]{}
Harris, D. E., Grindlay, J. E., 1979,
{\it MNRAS}, {\bf 188}, 25
\bibitem[]{}
Harris, D. E., Stern, C. P., 1987,
{\it ApJ}, {\bf 313}, 136
\bibitem[]{}
Harris, D. E., Carilli, C. L., Perley, R. A., 1994,
{\it Nature}, {\bf 367}, 713
\bibitem[]{}
Harris, D. E., Leighly, K. M., Leahy, J. P., 1998,
{\it ApJ}, {\bf 499}, L 149
\bibitem[]{}
Harris, D. E., Hjorth, J., Sadun, A. C., Silverman, J. D., 
Vestergaard, M., 1999, 
{\it ApJ}, {\bf 518}, 213
\bibitem[]{}
Harris, D. E., Nulsen, P. E. J., Ponman, T. J., et al., 2000, 
{\it ApJ}, {\bf 530}, L 81
\bibitem[]{}
Harris, D. E., Krawczynski, H., 2002,
{\it ApJ}, {\bf 565}, 244
\bibitem[]{}
Heavens, A.F., Meisenheimer, K., 1987, {\it MNRAS}, {\bf 225}, 
335
\bibitem[]{}
Henry, J. P., Henriksen, M. J., 1986, {\it ApJ},
{\bf 301}, 689
\bibitem[]{}
Jaffe, W. J., 1977, {\it ApJ}, {\bf 216}, 212
\bibitem[]{}
Jester, S., R\"oser, H.-J., Meisenheimer, K.,
Perley, R.,  Conway, R., 2001, {\it A\&A}, {\bf 373}, 447
\bibitem[]{}
Jetser, S., R\"oser, H.-J., Meisenheimer, K.,
Perley, R., 2002, {\it A\&A}, {\bf 385}, L27
\bibitem[]{}
Jones, T. W., O'Dell, S. L., Stein, W. A., 1974a,
{\it ApJ}, {\bf 188}, 253
\bibitem[]{}
Jones, T. W., O'Dell, S. L., Stein, W. A., 1974b,
{\it ApJ}, {\bf 192}, 259
\bibitem[]{}
Isenberg, P. A., 1987, {\it JGR}, {\bf 92}, 1067
\bibitem[]{}
Kaneda, H., Tashiro, M., Ikebe, Y., et al., 1995,
{\it ApJ}, {\bf 453}, L13
\bibitem[]{}
Kardashev, N. S., 1962, {\it Soviet Astronomy - AJ},
{\bf 6}, 3
\bibitem[]{}
Katz--Stone, D. M., Rudnick L., 1997, {\it ApJ}, {\bf 488},
146
\bibitem[]{}
Katz--Stone, D. M., Rudnick L., Butenhoff, C., O'Donoghue, A. A.,
1999, {\it ApJ}, {\bf 516}, 716
\bibitem[]{}
Keel, W. C., 1988, {\it ApJ}, {\bf 329}, 532
\bibitem[]{}
Kirk, J. G., Rieger, F. M., Mastichiadis, A., 1998,
{\it A\&A}, {\bf 333}, 452
\bibitem[]{}
Kirk, J. G., Guthmann, A. W., Gallant, Y. A., Achterberg, A.,
2000, {\it ApJ}, {\bf 542}, 235
\bibitem[]{}
Komissarov, S. S., 1990, {\it Sov. Astronomy Lett.}, {\bf 16},
284
\bibitem[]{}
Kraft R. P., Forman W. R., Jones C., et al., 2002, {\it ApJ},
{\bf 569}, 54
\bibitem[]{}
Kundt, W., Gopal-Krishna, 1980, {\it Nature}, {\bf 288}, 
149
\bibitem[]{}
Lacombe, C., 1977, {\it A\&A}, {\bf 54}, 1
\bibitem[]{}
L\"ahteenm\"aki, A., Valtaoja, E., 1999, 
{\it ApJ}, {\bf 521}, L493
\bibitem[]{}
Laing, R. A., 1993, in {\it Astrophysical Jets}, D.Burgarella,
M.Livio,C.P.O'Dea eds. (Cambridge University Press, Cambridge),
p.95
\bibitem[]{}
Laing, R. A., 1986, in {\it Energy transport in radio galaxies and
quasars, ASP Series}, {\bf 100}, 241
\bibitem[]{}
Laing, R. A., Parma, P., de Ruiter, H. R., Fanti, R., 
1999, {\it MNRAS}, {\bf 306}, 513
\bibitem[]{}
Leahy, J. P., 1991, in {\it Beams and Jets in Astrophysics}, eds.
P.A.Hughes (Cambridge
astrophysics series), p.100
\bibitem[]{}
Livshitz, E. M., Pitaevskii, L. P., 1981, {\it Physical Kinetics}
(Pergamon: Oxford)
\bibitem[]{}
Marcowith, A., Kirk, J. G., 1999, {\it A\&A}, {\bf 347},
391
\bibitem[]{}
Marshall, H. L., Harris, D. E., Grimes, J. P., 
et al., 2001, {\it ApJ}, {\bf 549} L167
\bibitem[]{}
Marshall, H. L., Miller, B. P., Davis, D. S., et al., 2002,
{\it ApJ}, {\bf 564}, 683
\bibitem[]{}
Meisenheimer, K., Roser, H.-J., 1986, {\it Nature},
{\bf 319}, 459
\bibitem[]{}
Meisenheimer, K., Roser, H.-J., Hiltner, P. R., et al., 
1989, {\it A\&A}, {\bf 219}, 63
\bibitem[]{}
Meisenheimer, K., Yates, M. G., Roeser, H.-J., 1997, 
{\it A\&A}, {\bf 325}, 57
\bibitem[]{}
Meisenheimer, K., Haas, M., M\"uller, S. A. H.,
Chini, R., Klaas, U., Lemke, D., 2001, 
{\it A\&A}, {\bf 372}, 719
\bibitem[]{}
Melrose, D. B., 1980, {\it Plasma Astrophysics, Nonthermal Processes
in Diffuse Magnetized Plasmas} (Gordon \& Breach Science Publishers)
\bibitem[]{}
Micono, M., Zurlo, N., Massaglia, S., Ferrari, A.,
Melrose, D. B., 1999, {\it A\&A}, {\bf 349}, 323
\bibitem[]{}
Miley, G., 1980, {\it ARA\&A}, {\bf 18}, 165
\bibitem[]{}
Myers, S. T., Spangler, S. R., 1985,
{\it ApJ}, {\bf 291}, 52
\bibitem[]{}
Mushotzky, R. F., Scharf, C. A., 1997, {\it ApJ}, 
{\bf 482}, L13
\bibitem[]{}
Neumann, D. M., 1999, {\it ApJ}, {\bf 520}, 87
\bibitem[]{}
Ostrowski, M., 2000, {\it MNRAS}, {\bf 312}, 579
\bibitem[]{}
Pacholczyk, A. G., 1970, {\it Radio Astrophysics}
(W.H. Fereeman and Company: San Francisco)
\bibitem[]{}
Pesce, J. E., Sambruna R. M., Tavecchio, F., et al., 2001,
{\it ApJ}, {\bf 556}, L79
\bibitem[]{}
Petrosian, V., 2001, {\it ApJ}, {\bf 557}, 560
\bibitem[]{}
Prieto, M. A., 1997, {\it MNRAS}, {\bf 284}, 627
\bibitem[]{}
Rawlings, S., Saunders, R., 1991, {\it Nature}, {\bf 349}, 
138
\bibitem[]{}
Rees, M. J., 1978, {\it Nature}, {\bf 275}, 35
\bibitem[]{}
Rybicki, G. B., Lightman A. P., 1979, {\it Radiative Processes in
Astrophysics} (Wiley, New York)
\bibitem[]{}
Sambruna, R. M., Urry, C. M., Tavecchio, F., 2001, 
{\it ApJ}, {\bf 549}, L161
\bibitem[]{}
Sambruna, R. M., Maraschi, L., Tavecchio, F., et al.,
2002, {\it ApJ}, {\bf 571}, 206
\bibitem[]{}
Sanders, D. B., Phinney, E. S., Neugebauer, G., Soifer, B. T., Matthews, 
K., 1989, {\it ApJ}, {\bf 347}, 29
\bibitem[]{}
Sarazin, C. L., 1999, {\it ApJ}, {\bf 520}, 529
\bibitem[]{}
Saslaw, W. C., Tyson, J. A., Crane, P., 1978, 
{\it ApJ}, {\bf 222}, 435
\bibitem[]{}
Schlickeiser, R., 1996, {\it A\&AS}, {\bf 120}, 481
\bibitem[]{}
Schwartz, D. A., Marshall, H. L., Lovell, J. E. J., et al., 
2000, {\it ApJ}, {\bf 540}, L69
\bibitem[]{}
Siemiginowska, A., Bechtold, J., Aldcroft, T. L.,
et al., 2002, {\it ApJ}, {\bf 570}, 543
\bibitem[]{}
Simkin, S. M., 1978, {\it ApJ}, {\bf 222}, L55
\bibitem[]{}
Stawarz, L., Ostrowski, M., 2002a, {\it PASA}, {\bf 19}, 22
\bibitem[]{}
Stawarz, L., Ostrowski, M., 2002b, {\it ApJ} in press;
astro-ph/0203040
\bibitem[]{}
Tashiro, M., Kaneda, H., Makishima, K., et al., 1998,
{\it ApJ}, {\bf 499}, 713
\bibitem[]{}
Tashiro, M., Makishima, K., Kaneda, H., 2000, 
{\it AdSpR}, {\bf 25}, 751
\bibitem[]{}
Tashiro, M., Makishima, K., Iyomoto, N., Isobe, N., Kaneda, H., 
2001, {\it ApJ}, {\bf 546}, L19
\bibitem[]{}
Tavecchio, F., Maraschi, L., Sambruna, R. M., Urry, C. M., 
2000, {\it ApJ}, {\bf 544}, L23
\bibitem[]{}
Thomson, R. C., Crane, P., Mackay, C. D., 1995, 
{\it ApJ}, {\bf 446}, L93
\bibitem[]{}
Tribble, P. C., 1993, {\it MNRAS}, {\bf 263}, 31
\bibitem[]{}
Verschur, G. L., Kellermann, K. I., 1988, 
{\it Galactic and Extragalactic Radio Astronomy}
(Springer--Verlag)
\bibitem[]{}
Webb, G. M., Drury, L. O'C., Biermann, P., 1984,
{\it A\&A}, {\bf 137}, 185
\bibitem[]{}
Webb, G. M., Fritz, K. D., 1990, {\it ApJ}, {\bf 362}, 
419
\bibitem[]{}
Wilson, A. S., Young, A. J., Shopbell, P. L., 2000, {\it ApJ},
{\bf 544}, L27
\bibitem[]{}
Wilson, A. S., Young, A. J., Shopbell, P. L., 2001, {\it ApJ},
{\bf 547}, 740
\bibitem[]{}
Wilson, A. S., Yang Y., 2002, {\it ApJ}, {\bf 568}, 133
\bibitem[]{}
Worrall, D. M., Birkinshaw, M., Hardcastle, M. J., 2001, 
{\it MNRAS}, {\bf 326}, L 7
\end{thebibliography}
\end{document}